\documentclass[fleqn,usenatbib]{mnras}

\usepackage{newtxtext,newtxmath}

\usepackage[T1]{fontenc}

\DeclareRobustCommand{\VAN}[3]{#2}
\let\VANthebibliography\thebibliography
\def\thebibliography{\DeclareRobustCommand{\VAN}[3]{##3}\VANthebibliography}

\usepackage{graphicx}	
\usepackage{amsmath}	

\usepackage{siunitx}
\usepackage{makecell}

\usepackage[utf8]{inputenc}

\newcommand{\app}[1]{Appendix~\ref{sec:#1}}

\newcommand{\fig}[1]{Figure~\ref{fig:#1}}
\renewcommand{\sec}[1]{Section~\ref{sec:#1}}
\newcommand{\tab}[1]{Table~\ref{tab:#1}}

\newcommand{\omegam}{$\Omega_{\mathrm{m}}$}
\newcommand{\sigmae}{$\sigma_{8}$}
\newcommand{\asnone}{$A_{\mathrm{SN1}}$}
\newcommand{\asntwo}{$A_{\mathrm{SN2}}$}
\newcommand{\aagnone}{$A_{\mathrm{AGN1}}$}
\newcommand{\aagntwo}{$A_{\mathrm{AGN2}}$}

\DeclareSIUnit\angstrom{\text {Å}}

\title[LtU: Parameter inference on photometry]{Learning the Universe: Cosmological and Astrophysical Parameter Inference with Galaxy Luminosity Functions and Colours}

\author[C. C. Lovell et al.]{Christopher C. Lovell,$^{1,2,3}$\thanks{E-mail: chris.lovell.astro@gmail.com (CCL)}
Tjitske Starkenburg$^{4,5,6}$,
Matthew Ho,$^{7}$
Daniel Angl\'{e}s-Alc\'{a}zar,$^{8}$
\newauthor
Romeel Dav\'{e},$^{9,10,11}$
Austen Gabrielpillai,$^{12}$
Kartheik G. Iyer,$^{7}$
Alice E. Matthews,$^{13}$
William J. Roper,$^{14}$
\newauthor
Rachel S. Somerville,$^{15}$
Laura Sommovigo,$^{15}$
Francisco Villaescusa-Navarro$^{15,16}$
\\
\\
$^{1}$Kavli Institute for Cosmology, Madingley Road, Cambridge CB3 0HA, UK\\
$^{2}$Institute of Astronomy, Madingley Road, Cambridge CB3 0HA, GB\\
$^{3}$Institute of Cosmology and Gravitation, University of Portsmouth, Burnaby Road, Portsmouth, PO1 3FX, UK\\
$^{4}$Center for Interdisciplinary Exploration and Research in Astrophysics (CIERA), Northwestern University, 1800 Sherman Ave, Evanston IL 60201, USA\\
$^{5}$Department of Physics and Astronomy, Northwestern University, 2145 Sheridan Rd, Evanston IL 60208, USA\\
$^{6}$NSF-Simons AI Institute for the Sky (SkAI), 172 E. Chestnut St., Chicago, IL 60611, USA\\
$^{7}$Columbia Astrophysics Laboratory, Columbia University, 550 West 120th Street, New York, NY 10027, USA\\
$^{8}$Department of Physics, University of Connecticut, 196 Auditorium Road, U-3046, Storrs, CT 06269-3046, USA\\
$^{9}$Institute for Astronomy, Royal Observatory, University of Edinburgh, Edinburgh EH9 3HJ, UK\\
$^{10}$University of the Western Cape, Department of Physics and Astronomy, Bellville, Cape Town 7535, South Africa\\
$^{11}$South African Astronomical Observatories, Observatory, Cape Town 7925, South Africa\\
$^{12}$Department of Astrophysics, The Graduate Center, City University of New York, 365 5th Ave, New York, NY 10016, USA\\
$^{13}$Mullard Space Science Laboratory, University College London, Holmbury St Mary, Dorking, Surrey, RH5 6NT, UK\\
$^{14}$Astronomy Centre, University of Sussex, Falmer, Brighton BN1 9QH, UK\\
$^{15}$Center for Computational Astrophysics, Flatiron Institute, 162 Fifth Avenue, New York, NY, 10010, USA\\
$^{16}$Department of Astrophysical Sciences, Princeton University, Princeton NJ 08544, USA
}

\date{Accepted XXX. Received YYY; in original form ZZZ}

\pubyear{\the\year{}}

\begin{document}
\label{firstpage}
\pagerange{\pageref{firstpage}--\pageref{lastpage}}
\maketitle

\begin{abstract}
We perform the first direct cosmological and astrophysical parameter inference from the combination of galaxy luminosity functions and colours using a simulation based inference approach.
Using the \textsc{Synthesizer} code we simulate the dust attenuated ultraviolet--near infrared stellar emission from galaxies in thousands of cosmological hydrodynamic simulations from the CAMELS suite, including the \textsc{Swift-EAGLE}, \textsc{IllustrisTNG}, \textsc{Simba} \& \textsc{Astrid} galaxy formation models.
For each galaxy we calculate the rest-frame luminosity in a number of photometric bands, including the SDSS \textit{ugriz} and GALEX FUV \& NUV filters; this dataset represents the largest catalogue of synthetic photometry based on hydrodynamic galaxy formation simulations produced to date, totalling >200 million sources.
From these we compile luminosity functions and colour distributions, and find clear dependencies on both cosmology and feedback.
We then perform simulation based (likelihood-free) inference using these distributions to obtain constraints on \omegam, \sigmae, and four parameters controlling the strength of stellar and AGN feedback.
Both colour distributions and luminosity functions provide complementary information on certain parameters when performing inference.
We achieve constraints on the stellar feedback parameters, as well as \omegam and \sigmae.
The latter is attributable to the fact that the photometry encodes the star formation--metal enrichment history of each galaxy; galaxies in a universe with a higher \sigmae\ tend to form earlier and have higher metallicities, which leads to redder colours.
We find that a model trained on one galaxy formation simulation generalises poorly when applied to another, and attribute this to differences in the subgrid prescriptions, and lack of flexibility in our emission modelling.
The photometric catalogues are publicly available\footnotemark.
\end{abstract}

\begin{keywords}
galaxies: abundances -- galaxies: photometry -- cosmology: cosmological parameters
\end{keywords}


\footnotetext{\url{https://camels.readthedocs.io/en/latest/photometry.html}}

\section{Introduction}
\label{sec:intro}

Machine learning methods, particularly deep learning algorithms, are now capable of learning both deterministic and probabilistic relationships with a high degree of flexibility, and have achieved wide use within the fields of cosmology and astrophysics \citep{carleo_machine_2019,huertas-company_dawes_2023,smith_astronomia_2023}.
One particularly powerful application of these methods is in Bayesian inference, wherein one estimates the parameters of a model given some assumed prior hypothesis and observed data \citep[\textit{e.g.}][]{christensen_bayesian_2001}.
This combination of machine learning and Bayesian inference is known as Simulation Based Inference \citep[SBI, also called implicit likelihood inference, or likelihood-free inference; see][]{marin_approximate_2012,cranmer_frontier_2020}, a rigorous Bayesian approach to learning the relationship between simulated data ($\mathbf{X}$) and the underlying parameters used to generate it ($\theta$), typically leveraging flexible neural density estimators \citep[NDEs, e.g.][]{durkan_neural_2019}.
In contrast with traditional Bayesian methods, an analytic form of the likelihood does not need to be provided; instead, an SBI approach directly learns the form of the likelihood.
This provides many advantages, particularly in instances where the likelihood is non-Gaussian (such as when systematics are considered; \citealt{jeffrey_likelihood-free_2021}), or intractable (such as in field--level inference; \citealt{leclercq_accuracy_2021,lemos_robust_2023}).
They also permit amortized inference, whereby the full posterior can be evaluated and sampled rapidly at the point of inference, reducing the computational cost; this opens up new avenues for Bayesian parameter inference on extremely large datasets \citep{ho_ltu-ili_2024}.

A number of studies have used SBI to perform cosmological parameter inference.
These exploit simulations that explicitly model the relationship between cosmological parameters and observable statistics, such as the cosmic microwave background power spectrum \citep[e.g.][]{lemos_robust_2023}, or clustering statistics in the presence of weak-lensing effects \citep[e.g.][]{von_wietersheim-kramsta_kids-sbi_2025}.
These simulations cover a wide range of sophistication, from analytical descriptions of the distribution of halo properties \citep{mead_hmcode-2020_2021,tessore_glass_2023} to full $N$-body simulations of the large scale structure \citep{villaescusa-navarro_quijote_2020}.
Analytic or empirical models are often then applied to map galaxies or observational tracers to the underlying matter density.
However, these simple models relating dark matter properties to observational tracers are often inadequate \citep{hadzhiyska_limitations_2020}, and at small scales astrophysical effects, such as energetic feedback from active galactic nuclei (AGN) and star formation, can impact the overall matter distribution \citep{angles-alcazar_cosmic_2017,borrow_cosmological_2020,delgado_predicting_2023,tillman_exploration_2023,gebhardt_cosmological_2024}, requiring modifications to $N$-body predictions \citep{bose_road_2021,carrilho_road_2022,schaller_flamingo_2025}.
To overcome these limitations requires simulating these processes explicitly.

State-of-the-art hydrodynamic simulations are now capable of producing realistic galaxy samples in large volumes that match a number of key distribution functions in the local Universe \citep{somerville_physical_2015,vogelsberger_cosmological_2020,crain_hydrodynamical_2023}.
These simulations self-consistently model the evolution of collisionless dark matter as well as baryons down to some resolution limit, beyond which subgrid models are used to approximate the impact of sub-resolution physical processes on the macro-scale properties of galaxies.
Such simulations can be used in an SBI framework not only to understand the impact of baryons and astrophysical processes on cosmological parameter inference, but also to perform inference directly on astrophysical parameters of interest.
However, hydrodynamic simulations are exceedingly computationally expensive, limiting the volume, resolution and / or physical fidelity of the simulations that can be run.
Beyond the challenges inherent in running the simulations, they are also computationally challenging for use in SBI due to the range of parameters that can be varied, and the large volume of data products that must be computed and stored.

Despite these obstacles there has been recent progress in producing hydrodynamic simulation suites that are amenable to SBI approaches.
The Cosmology and Astrophysics with MachinE-Learning Simulations \citep[CAMELS;][]{villaescusa-navarro_camels_2021} project brought together a number of hydrodynamic and semi-analytic \citep{perez_constraining_2023} galaxy formation models, each consisting of large suites of simulations varying both cosmological and astrophysical paramaters that have been used in an SBI framework for parameter inference, both for posterior summaries \citep[][]{villaescusa-navarro_multifield_2021,villaescusa-navarro_cosmology_2022,nicola_breaking_2022} and direct neural density estimation of posteriors \citep{hassan_hiflow_2022,friedman_higlow_2022,thiele_percent-level_2022,villanueva-domingo_learning_2022,shao_robust_2023,de_santi_robust_2023,de_santi_field-level_2025,jo_calibrating_2023,echeverri-rojas_cosmology_2023}.

The majority of these studies have performed inference on intrinsic properties of galaxies (or physical fields).
However, in order to make a like-for-like comparison with observations we must forward model the electromagnetic emission from galaxies, and compare directly in this observer space \citep{fortuni_forecast_2023}.
There are a number of approaches for achieving this, but almost all start by modelling the stellar emission using stellar population synthesis (SPS) models \citep{conroy_modeling_2013} coupled to the star formation and metal enrichment history of each simulated galaxy.
Many also account for line and continuum emission from young nebular regions, utilising photoionisation codes such as \textsc{Mappings} and \textsc{Cloudy} \citep{dopita_spectral_1996,chatzikos_2023_2023}.
One of the most important galaxy components that has an outsized impact on the observed spectral energy distribution (SED) is cosmic dust, which leads to attenuation in the UV--optical, and thermal emission in the infrared \citep{calzetti_effects_2001,draine_interstellar_2003}.
There are a number of methods for modelling this attenuation and emission, from simple analytic models \citep{charlot_simple_2000} to full dust radiative transfer approaches \citep{camps_skirt_2015,narayanan_powderday_2021}.

It is clear from these various ingredients in the forward model that many degeneracies can be introduced.
For example, the choice of SPS model can have a large impact on the predicted intrinsic emission, particularly if the impact of binaries is taken into account \citep{wilkins_photometric_2016,stanway_stellar_2016,eldridge_binary_2017,tortorelli_impact_2024,bellstedt_progeny_2025}.
Photoionisation modelling introduces a large number of additional parameters, such as the hydrogen density of the incident cloud, the value of the assumed ionisation parameter, as well as the chemical abundances of the cloud and source \citep{byler_nebular_2017,wilkins_nebular-line_2020}.
The dust attenuation law can also take a number of forms, dependent on the grain size distribution and composition \citep{steinacker_three-dimensional_2013}, and depend on the intrinsic properties of galaxies in various ways \citep{salim_dust_2020,trayford_fade_2020,hahn_iq_2022}, such as by scaling the overall attenuation by the mass and metallicity of star forming gas \citep[see][]{trayford_colours_2015,lovell_learning_2019,vijayan_first_2021,pallottini_survey_2022}.
Despite this flexibility in the forward model, achieving a match to observations has still proven exceedingly difficult.
Optical luminosity functions, well constrained observationally in the local ($z \sim 0.1$) Universe, have been reproduced with varying success, and accurate colour distributions have proven to be even more elusive; achieving the right balance of blue (star-forming) to red (passive, dust obscured) galaxies is a complex challenge \citep{trayford_colours_2015,trayford_optical_2017,nelson_first_2018,lagos_far-ultraviolet_2019,donnari_star_2019,bravo_rest-frame_2020,dickey_iq_2021,trcka_uv_2022,gebek_many_2024}.

Despite these difficulties, a few recent works have forward modelled observed galaxy properties from hydrodynamic simulations and used these within an SBI framework. 
Recently, \cite{choustikov_inferring_2025} forward modelled JWST NIRCam photometry from the Sphinx simulations to estimate the ionising emissivity of high-redshift galaxies. 
Of particular relevance for this study is the work of \cite{hahn_cosmology_2024}, who showed how competitive cosmological constraints can be achieved using galaxy photometry alone calculated from the IllustrisTNG suite of CAMELS; they combined the posteriors from individual galaxies to achieve tighter constraints, complementing the work of \cite{villaescusa-navarro_cosmology_2022,echeverri-rojas_cosmology_2023,chawak_cosmology_2024} that performed parameter inference on single galaxies.
Inference on other forward modelled properties from CAMELS, such as diffuse gas emission and absorption, has also been demonstrated \citep{wadekar_sz_2023,butler_contreras_x-ray_2023,tillman_exploration_2023}.

In this paper we present the first simultaneous constraints of cosmological and astrophysical parameters using observable galaxy distribution functions, forward modelled from the CAMELS simulation suite.
All photometric catalogues are made publicly available at \url{https://camels.readthedocs.io/en/latest/data_access.html} for the community to use.
This paper is released as part of the Learning the Universe collaboration,\footnote{\url{https://learning-the-universe.org/}} which seeks to learn the cosmological parameters and initial conditions of the Universe by leveraging Bayesian forward modelling approaches.
The hope is that by better understanding the link between cosmology, astrophysics, and forward modelled observables, we can build better models for performing this kind of inference.
The paper is arranged as follows.
In \sec{camels} we describe the CAMELS simulations, including details on the simulation sets used in this analysis.
We describe our forward model for predicting galaxy emission and extracting rest-frame photometry in \sec{forward_model}.
In \sec{distribution_functions} we explore the distribution functions from the various CAMELS suites, and the impact of the cosmological and astrophysical parameters.
We present our results in \sec{sbi}, as well as introducing the Learning the Universe Implicit Likelihood Inference (LtU-ILI) framework for SBI.
We discuss our results in \sec{discussion} as well as the difficulties of performing SBI directly on observed relations.
We summarise our conclusions in \sec{conclusions}.

\section{The CAMELS Simulations}
\label{sec:camels}

\begin{table*}
	\centering
	\caption{A summary of the physical meaning of the feedback parameters in each of the simulation suites.}
	\begin{tabular}{ccccc}
		\hline
		Simulation & \asnone\ & \asntwo\ & \aagntwo\ & \aagntwo\ \\
		\hline
        Swift-EAGLE & \makecell{Galactic winds: \\ energy per SNII \\ event $[0.25-4.0]$} & \makecell{Galactic winds: metallicity  \\ dependence of the stellar \\ feedback fraction per unit \\ stellar mass $[0.5-2.0]$} & \makecell{Thermal BH feedback: \\ scaling of the Bondi accretion \\ rate $[0.25-4.0]$} & \makecell{Thermal BH feedback: \\ temperature jump of gas \\ particles $[0.5-2.0]$} \\[0.3cm]
        Astrid & \makecell{Galactic winds:\\ energy per unit \\ SFR $[0.25-4.00]$} & \makecell{Galactic winds: wind speed \\ $[0.50-2.00]$} & \makecell{Kinetic mode BH feedback:\\ energy per unit BH \\ accretion $[0.25-4.00]$} & \makecell{Thermal mode BH feedback: \\ energy per unit BH \\ accretion $[0.25-4.00]$} \\[0.3cm]
		Simba & \makecell{Galactic winds: \\ mass loading \\ $[0.25-4.00]$} & \makecell{Galactic winds: wind speed \\ $[0.50-2.00]$} & \makecell{QSO \& jet mode BH feedback:\\ momentum flux $[0.25-4.00]$} & \makecell{Jet mode BH feedback: jet speed \\ $[0.50-2.00]$} \\[0.3cm]
		IllustrisTNG & \makecell{Galactic winds: \\ energy per unit \\ SFR $[0.25-4.00]$} & \makecell{Galactic winds: wind speed \\ $[0.50-2.00]$} & \makecell{Kinetic mode BH feedback: \\ energy per unit BH \\ accretion $[0.25-4.00]$} & \makecell{Kinetic mode BH feedback: \\ ejection speed / burstiness \\ $[0.50-2.00]$} \\
		\hline
	\end{tabular}
	\label{tab:parameters}
\end{table*}

The Cosmology and Astrophysics with MachinE Learning Simulations project \citep[CAMELS;][]{villaescusa-navarro_camels_2021} is a suite of over 14000 cosmological hydrodynamic and $N$-body simulations, designed to explore the effect of cosmological and astrophysical parameter choices on structure formation and galaxy evolution, particularly through their use within machine learning frameworks.
In this project we use the hydrodynamic simulations from four different codes and galaxy formation models:

\begin{enumerate}
    \item \textsc{Simba} \citep{dave_simba:_2019}, run with the \textsc{Gizmo} code \citep{hopkins_new_2015}.
    \item \textsc{IllustrisTNG} \citep{pillepich_simulating_2018,weinberger_supermassive_2018}, run with the \textsc{Arepo} code \citep{springel_e_2010,weinberger_arepo_2019}.
    \item \textsc{Astrid} \citep{bird_astrid_2022,ni_astrid_2022}, run with the \textsc{MP-Gadget} code. The addition of Astrid to the CAMELS suite is described in \citep{ni_camels_2023}, including details of modifications compared to the fiducial Astrid model.
    \item \textsc{Swift-EAGLE} \citep{schaye_eagle_2015,crain_eagle_2015}, run with the \textsc{Swift} code \citep{schaller_swift_2024} using the \textsc{Sphenix} smoothed particle hydrodynamics implementation \citep{borrow_sphenix_2022}. The addition of Swift-EAGLE to the CAMELS suite is described in Lovell et al. \textit{in prep.}.
\end{enumerate}

These models represent a combination of different subgrid models for gas cooling, chemical enrichment, star formation, stellar feedback, black hole assembly and AGN feedback.
There are also differences in the underlying hydrodynamics solvers; \textsc{Arepo} implements an unstructured Voronoi tesselation, \textsc{Gizmo} uses a meshless finite-mass method, and \textsc{MP-Gadget} and \textsc{Swift} utilise Smoothed Particle Hydrodynamics (SPH).
Together these differences lead to significant macro differences between the predicted galaxy properties \citep{villaescusa-navarro_camels_2021,ni_astrid_2022}, however the fiducial simulations all reproduce the galaxy stellar mass function (as well as other key scaling relations) at $z = 0$ to reasonable accuracy.

Each simulation contains 256$^3$ dark matter and baryonic (gas particle) resolution elements, of mass $m_{\rm DM} = 6.49 \times 10^7 \, (\Omega_{\mathrm{m}} - \Omega_{\mathrm{b}})/0.251\, h^{-1} \, M_{\odot}$ and $m_{\rm b} = 1.27 \times 10^7 \, h^{-1} \, M_{\odot}$, respectively, within a periodic volume of $(25 \; h^{-1})^3 \; \mathrm{Mpc^3}$.
The initial conditions for each simulation were generated using second order Lagrangian perturbation theory using \textsc{MUSIC} \citep{hahn_multi-scale_2011}, assuming the same initial power spectrum of gas and dark matter particles at $z = 127$.
In all simulations a number of parameters are kept fixed, such as $\Omega_{\mathrm{b}} = 0.049$, $h = 0.6711$, $n_s =
0.9624$, $M_{\nu} = 0.0 \; \mathrm{eV}$, $w = -1$, $\Omega_K = 0$.
The differences between simulations are in the initial random phases, as well as 4 astrophysical parameters ($A_{\mathrm{SN1}}$, $A_{\mathrm{SN2}}$, $A_{\mathrm{AGN1}}$, $A_{\mathrm{AGN2}}$) and 2 cosmological parameters ($\Omega_{\mathrm{m}}$, $\sigma_{8}$).
We stress that, due to differences in the subgrid model implementations, the astrophysical parameters in each suite have different meanings.
This is true even in the case where two simulations vary the same quantity (e.g. the wind speed), because of the precise implementation details, but also because different models hold different secondary parameters constant whilst varying this primary variable.
The physical meaning of each parameter in each galaxy formation model is described in \tab{parameters}.

Structures are first identified with the Friends-of-Friends algorithm \citep[FOF;][]{davis_evolution_1985} and substructrues with \textsc{Subfind} \citep{springel_populating_2001}.
We treat the \textsc{Subfind} outputs as discrete galaxies, where the total stellar mass is $> 10^8 \, \mathrm{M_{\odot}}$ (corresponding to at least 5 star particles).
For the CV sets this gives $1613\pm80$ galaxies in Swift-EAGLE, $1188\pm59$ for Astrid, $1495\pm67$ for Simba and $878\pm34$ for IllustrisTNG at $z = 0.1$. The median values for the LH sets are similar, but the variance is obviously much larger.
We note that resolution effects due to discrete sampling of the star formation history can have a large impact on the resulting emission, particularly in the UV \citep{trayford_colours_2015}.
We will explore methods for smoothing the recent star formation in future work.
In this work we extract the age and metallicity of each star particle in each subhalo, and model the emission from each star particle independently (described in \sec{forward_model}).

There are a number of different simulation sets within each CAMELS suite.
The Cosmic Variance (CV) set contains 27 simulations with identical (fiducial) parameters, only changing the initial random number seed, in order to provide a means of assessing the impact of cosmic variance. 
The Latin Hypercube (LH) set contains 1000 simulations where the 6 parameters are varied using a latin hypercube within the following ranges: \omegam\ $\in [0.1, 0.5]$, \sigmae\ $\in [0.6, 1.0]$, \asnone\ $\in [0.25, 4.0]$, \asntwo\ $\in [0.5, 2.0]$, \aagnone\ $\in [0.25, 4.0]$, and \aagntwo\ $\in [0.5, 2.0]$\footnote{For Astrid, the \aagntwo\ parameter also ranges from 0.25 to 4.0}.
The 1 Parameter (1P) set contains 25 simulations where one parameter at a time is varied around the fiducial parameters.
In the 1P and CV sets the fiducial cosmological parameters are $\Omega_{\mathrm{m}} = 0.3$ and $\sigma_{8} = 0.8$, and the fiducial astrophysical parameters are defined at $A = 1.0$.

The full CAMELS dataset is available online at \url{https://camels.readthedocs.io/en/latest/data_access.html}, and described in \cite{villaescusa-navarro_camels_2023}.
We generate photometry for galaxies in all sets (pipeline described in \sec{forward_model}, and catalogue in \app{photo_database}), and these are also made available online at the same address.
\begin{figure*}
	\includegraphics[width=\textwidth]{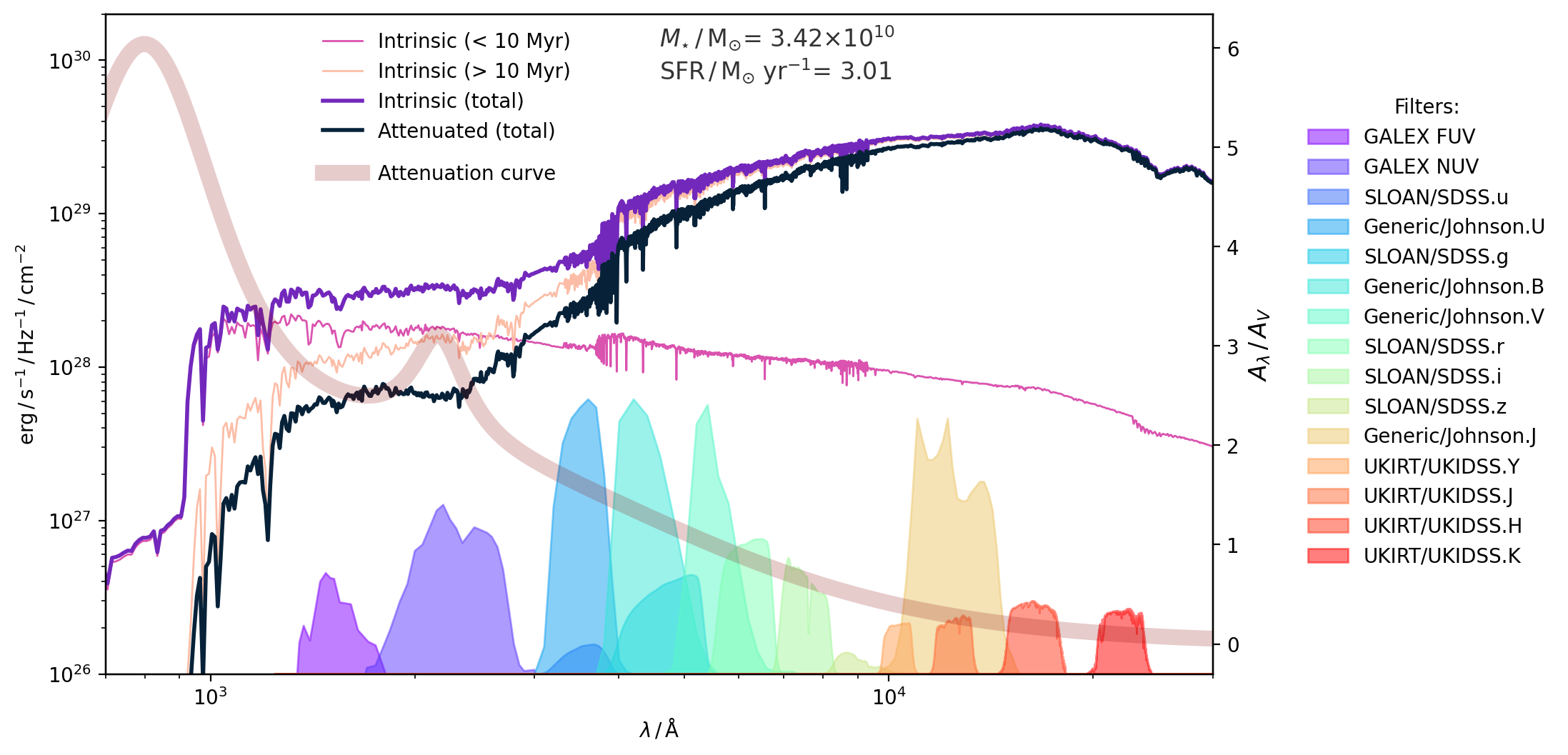}
    \caption{Spectral energy distribution of an example galaxy from the \textsc{IllustrisTNG} CV set at $z = 0.1$.
	Shown is the intrinsic emission from young and old stars, the combined intrinsic emission, as well as the total attenuated emission.
	Filter transmission curves for some of the key rest-frame filters used in this work are also shown.}
    \label{fig:spectra_example}
\end{figure*}

\section{Forward modelling with Synthesizer}
\label{sec:forward_model}

In order to model the emission for all galaxies in the CAMELS simulation suites we employ a forward model for the UV-optical spectral energy distribution that takes into account stellar emission and the impact of dust attenuation.
Importantly, we choose a forward model that is agnostic to the specific subgrid implementation in each galaxy formation model, whether that is \textsc{IllustrisTNG}, \textsc{Swift-EAGLE}, \textsc{Astrid} or \textsc{Simba}.
This ensures that the results for each model are comparable; the emission of each galaxy is dependent only on its star formation and metal enrichment history, itself derived from the distribution of star particle properties.

We use \textsc{Synthesizer}\footnote{\url{https://synthesizer-project.github.io/}} \citep{lovell_synthesizer_2025,roper_synthesizer_2025} to produce synthetic observables, leveraging flexibility and computational efficiency.
Below we describe the different ingredients that go into our \textsc{Synthesizer} forward model, and the resulting luminosity and colour distributions.

\subsection{Stellar population modelling}

The intrinsic stellar emission is produced by coupling each star particle in a galaxy to a stellar population synthesis (SPS) model, based on its age, metallicity and initial stellar mass\footnote{Initial stellar masses are not available for Astrid, so we assume that all star particles have an initial mass equal to 1/4 the initial gas element mass \citep{bird_astrid_2022}.}.
The `current' stellar mass of a star particle takes into account the recycling of mass due to evolved stellar populations; since SPS models implicitly account for this as a function of age, the initial stellar mass must be used as input.
The integrated emission of a galaxy is then the sum of the particle contributions.
In order to provide a qualitative assessment of the impact of the choice of SPS model, we generate the emission using two different SPS models: BC03 \citep{bruzual_stellar_2003} and BPASS \citep{eldridge_binary_2017,stanway_re-evaluating_2018}.
For BC03, we use the Padova 2000 tracks, described in \cite{girardi_evolutionary_2000,girardi_theoretical_2002}.
For BPASS we use v3.2, described in \cite{byrne_dependence_2022}.
For both models we assume a \cite{chabrier_galactic_2003} initial mass function (IMF), with a high mass cut off of 100 $\mathrm{M_\odot}$, and a low mass cut off of 0.1 $\mathrm{M_\odot}$.
This is identical to the IMF assumed in each galaxy formation model, which ensures consistency between the feedback and enrichment subgrid prescriptions and the forward modelled emission.

\cite{trayford_colours_2015} showed that applying a 3D or 2D aperture can impact the luminosities of extended, high mass objects, where up to 40\% of the light comes from an extended intracluster component.
However, we choose to neglect aperture corrections, for three reasons.
Firstly, the low resolution of the fiducial CAMELS simulations, in terms of mass ($\sim10^7 \; \mathrm{M_{\odot}}$ per resolution element) and force resolution, leads to poorly estimated galaxy sizes.
Secondly, since the volume of each simulation is relatively small, we do not simulate any massive clusters or groups; as such, we are not in a mass regime where aperture corrections are important \citep[but see][]{pillepich_simulating_2018}.
And thirdly, the precise aperture corrections will be strongly dependent on the exact survey configuration and redshift distribution of sources, a detail we do not consider in this work.
As a result, we choose not to apply any 2D or 3D apertures when estimating photometry, and instead select all bound star particles within the subhalo. 

We do not model the impact of nebular emission from young star forming regions, nor the contribution of active galactic nuclei; we leave an assessment of the impact of these components on our results to future work.
We also ignore the effect of the clumpy intergalactic medium (IGM), since we concentrate on rest-frame bands redward of Lyman-$\alpha$, and since this effect will not have a large impact below $z \sim 3$.

We adopt the BC03 models as our fiducial SPS model throughout the manuscript, unless otherwise stated.
\fig{spectra_example} shows the integrated intrinsic emission in the UV-optical wavelength range from an example galaxy taken from one of the IllustrisTNG CV set simulations.

\subsection{Dust model}
\label{sec:dust_model}

Dust constitutes only around 0.1\% of the total baryons in the Universe today, but has an outsize effect on galaxy emission; around 30\% of all photons today have been reprocessed by dust grains \citep{bernstein_first_2002}.
To account for the effect of dust in a way that is agnostic to the underlying galaxy formation model, we adopt a simple two-component dust model that differentially attenuates young and old stars, analogous to that first presented by \cite{charlot_simple_2000}.
The model varies the dust optical depth $\tau$ as follows,
\begin{align}
    &\tau = \tau_{\rm cloud} + \tau_{\rm ISM}  &t \leqslant t_{\rm disp}\\
    &\tau = \tau_{\rm ISM} &t > t_{\rm disp}
\end{align}
where $\tau_{\mathrm{cloud}} = 0.67$, $\tau_{\mathrm{ISM}}= 0.33$\footnote{Assuming a hydrogen density $N_{\rm H}^{\rm cloud} = 1.37 \times 10^{21} \, \mathrm{cm^{-2}}$ and $N_{\rm H}^{\rm ISM} = 6.79 \times 10^{20} \, \mathrm{cm^{-2}}$, for an extinction cross section $C_{\rm ext}^{V} = 4.86 \times 10^{-22}$ \citep{draine_interstellar_2003}, all calculated at visible wavelengths, $\lambda_V = \SI{5500}{\angstrom}$.} and the dispersion time $t_{\mathrm{disp}} = 10 \; \mathrm{Myr}$.
These parameters were calibrated to nearby galaxies, and have been used in previous studies employing similar models \citep{genel_introducing_2014,trayford_colours_2015}.
The only property of each galaxy (and by extension each galaxy formation model) that directly affects the level of attenuation is the ratio of stellar mass formed before or after $t_{\mathrm{disp}}$.

We use a fixed Milky Way attenuation curve \citep{pei_interstellar_1992}, parametrised using the analytic form of \cite{li_dust_2008}.
The transmission at wavelength $\lambda$ for a particle with age $t$ is then given by
\begin{align}
    T(\lambda, t) = \mathrm{exp} \left[ - \left( \dfrac{ \tau_{\rm MW}(\lambda) }{ \tau_{\rm MW}( \lambda = \SI{5500}{\angstrom} ) } \right) \times ( \tau_{\rm cloud} + \tau_{\rm ISM} ) \right]
\end{align}
We do not model thermal re-emission from dust, nor scattering; the latter has been shown to contribute negligibly to the galaxy attenuation curve \citep{fischera_starburst_2003}.

In Simba, the properties of dust are predicted self-consistently using an on-the-fly creation, growth and destruction model \citep{li_dust--gas_2019}; in the interests of consistency with the other galaxy formation models, we ignore these properties, and adopt the model above. 
In future work we will explore the impact of our assumed dust model, and perform direct inference on these dust parameters, as well as marginalisation over the effect of dust in cosmological inference.
We will also explore how to link the attenuation to the properties of the galaxy (e.g. the metallicity of star-forming gas) in a model agnostic way.

\subsection{Photometry}
\label{sec:dist_funcs}

We generate rest-frame photometry from each of our intrinsic and dust--attenuated galaxy SEDs in a range of photometric filters.
These include the GALEX FUV \& NUV bands, top-hat filters centred at $\SI{1500}{\angstrom}$ and $\SI{2800}{\angstrom}$, SDSS \textit{ugriz}, UKIRT UKIDSS \textit{YJHK}, and Johnson \textit{UBVJ} bands\footnote{Transmission curves obtained from the Spanish Virtual Observatory Filter Profile Service,  \citep{rodrigo_svo_2012,rodrigo_svo_2020}, \url{http://svo2.cab.inta-csic.es/theory/fps/index.php?mode=voservice}.}.
These filter profiles are shown in \fig{spectra_example}.
We additionally calculate rest-frame and observer-frame fluxes in the HST ACS F435W, F606W, F775W, F814W, and F850LP bands, the HST WFC3 F098M, F105W, F110W, F125W, F140W and F160W, and JWST NIRCam F070W, F090W, F115W, F150W, F200W, F277W, F356W and F444W.
We do not use these bands in this study, but make them publicly available for the community.
Formally, the luminosity $L$ in each band $X$ is defined as so,
\begin{align}
	L_X \equiv \frac{\int_{0}^{\infty} L_{\nu} \, S_{X}(\nu) \, d\nu}{\int_{0}^{\infty} S_{X}(\nu) \, d\nu} \;,
\end{align}
where $S_{X}(\nu)$ is the transmission at a given frequency for band $X$.
We then convert the calculated luminosities to absolute AB magnitudes,
\begin{align}
	M_X = -2.5 \; \mathrm{log_{10}} \left(\frac{L_{\nu} \,/\, (\mathrm{erg \,/\, s \,/\, Hz})}{d = 10 \mathrm{pc}} \right) - 48.6	\;\;,
\end{align}
where $d$ is the distance modulus and $L_{\nu}$ is the luminosity density.

We do this for all four galaxy formation models over 34 snapshots in the redshift range $z \in [0, 6]$.
The catalogues for each simulation are publicly available at \url{https://camels.readthedocs.io/}.
For this study we use these magnitudes to build rest-frame luminosity functions and colour distributions at a range of redshifts, explored in more detail below.

\begin{figure*}
	\includegraphics[width=\textwidth]{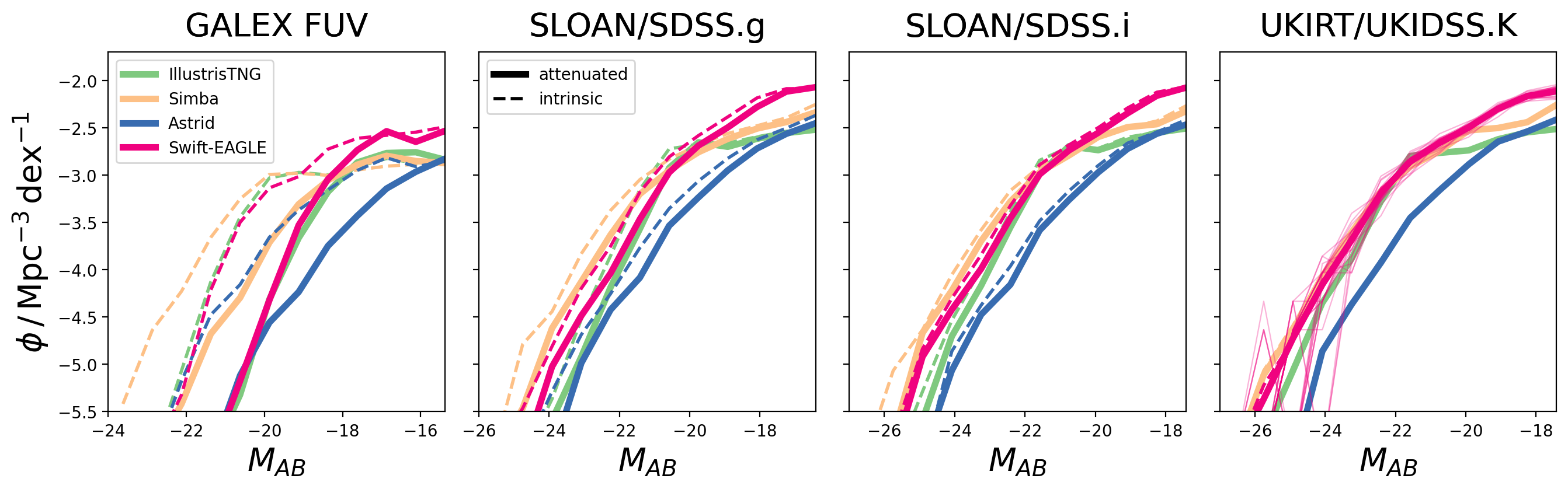}
    \caption{CV set Luminosity functions (LFs; in AB magnitudes) at $z = 0.1$ for the \textsc{IllustrisTNG}, \textsc{Simba}, \textsc{Astrid} and \textsc{Swift-EAGLE} simulation suites in the GALEX FUV, SDSS $g$ and $i$, and UKIRT K bands. Each LF is built using galaxies from all CV simulations combined.
	Each panel shows the LFs obtained from the attenuated (solid lines) and intrinsic (dashed lines) emission.
    For Swift-EAGLE in the UKIRT K band we additionally show each individual CV simulation as thin red line, to illustrate the impact of cosmic variance.
    }
    \label{fig:CV_sims_lfs}
\end{figure*}

\begin{figure*}
	\includegraphics[width=\textwidth]{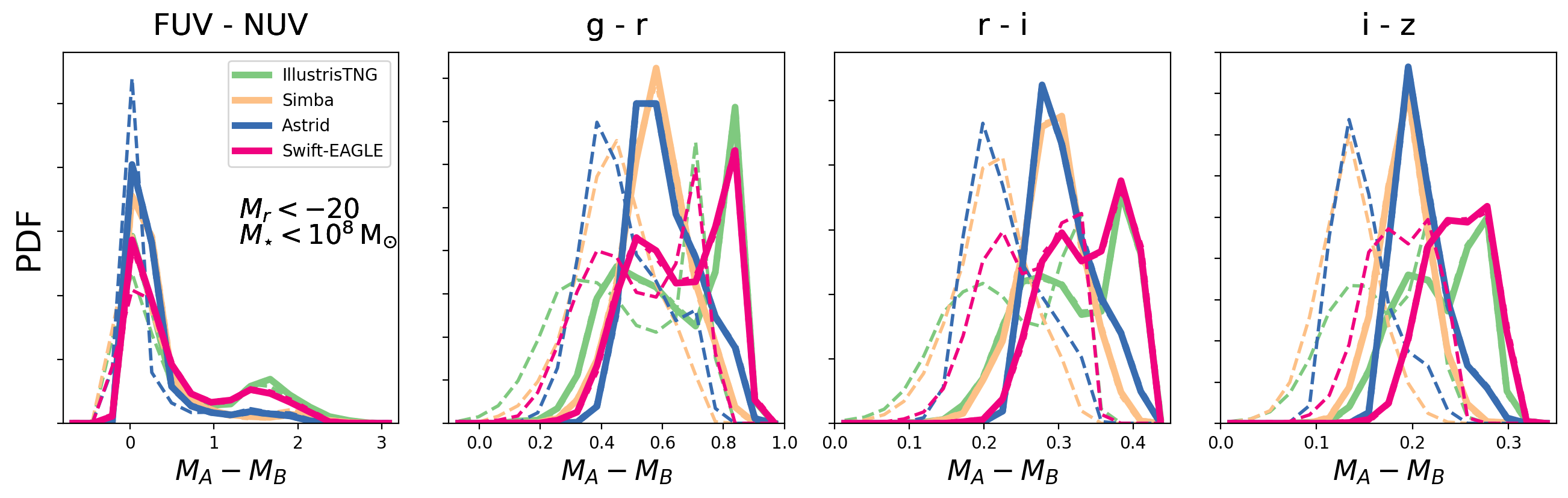}
    \caption{The same as \fig{CV_sims_lfs}, but showing normalised colour distributions. An additional magnitude cut for galaxies above $M_r < -20$ has been applied to remove faint galaxies. Left to right: GALEX $\mathrm{FUV}-\mathrm{NUV}$, SDSS $g-r$, $r-i$, $i-z$.}
    \label{fig:CV_sims_colours}
\end{figure*}

\section{Luminosity Functions and Colours in CAMELS}
\label{sec:distribution_functions}

\subsection{CV set distributions}

\begin{figure*}
	\includegraphics[width=\textwidth]{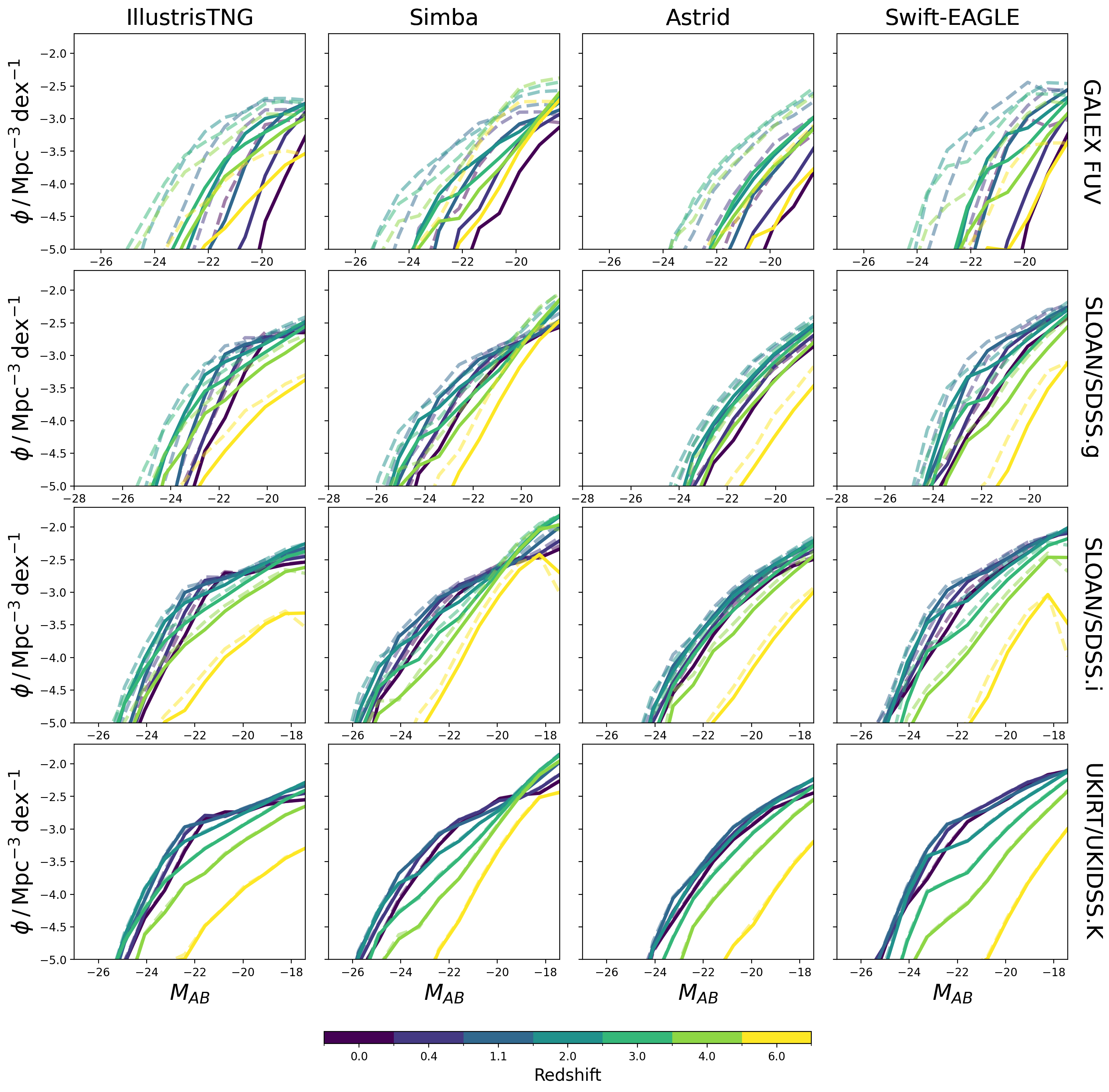}
    \caption{Luminosity functions (LFs; in AB magnitudes) in the GALEX FUV, SDSS $g$ and $i$, and UKIRT K bands, over a range of redshifts ($z \in [0, 6]$). Each LF is built using galaxies from all CV simulations combined, normalised by the total volume ($27 \times (25 \, h^{-1})^3 \; \mathrm{Mpc^3}$). We show the relations for the \textsc{IllustrisTNG}, \textsc{Simba}, \textsc{Astrid} and \textsc{Swift-EAGLE} simulation suites.
	Each panel shows the LFs for both attenuated (solid lines) and intrinsic (faded dashed lines) emission.
    }
    \label{fig:CV_lfs}
\end{figure*}

\subsubsection{Galaxy formation model comparison}

The Cosmic Variance (CV) set contains 27 simulations, each using the same fiducial parameters, but varying the random seed of the initial conditions.
\fig{CV_sims_lfs} shows the rest-frame luminosity function (LF) for both central and satellite subhalos from all 27 CV set simulations combined, for each galaxy formation model at $z = 0.1$.
Each LF is normalised by the total combined volume of the simulations ($27 \times (25 \, h^{-1})^3 \; \mathrm{Mpc^3}$).
We show the GALEX FUV, SDSS $g$ and $i$ and 2MASS $K$ band to demonstrate the full wavelength range probed ($\SI{1500}{\angstrom}$ - $\SI{20000}{\angstrom}$).
We also show both the intrinsic and dust attenuated LFs in each case.
It's clear that there are significant differences between the different galaxy formation models in all bands.
Astrid has the lowest normalisation across the magnitude range in all bands, which matches what is seen for the galaxy stellar mass function \citep{ni_astrid_2022}.
Simba extends to brighter magnitudes in all bands, but most noticeably at the blue end, where there is almost an order of magnitude more $M_{\mathrm{FUV}} \sim 21$ galaxies than the other models.
This may reflect the slightly higher normalisation of the galaxy stellas mass function at the high mass end in Simba \citep{dave_simba:_2019}.
Both IllustrisTNG and Swift-EAGLE show similar behaviour in the form and normalisation at all wavelengths, but the latter tends to have more faint galaxies in all bands, by approximately 0.5 dex.
Dust attenuation has a more significant impact on bluer bands, as expected from the form of the attenuation curve.

We also show a number of UV-optical (normalised) colour distributions in \fig{CV_sims_colours}, including GALEX $\mathrm{FUV}-\mathrm{NUV}$ and SDSS $g-r$, $r-i$ and $i-z$.
These include all galaxies above our fiducial stellar mass limit of $10^{8} \; \mathrm{M_{\odot}}$ (centrals and satellites), as well as a further cut to remove faint galaxies below an $r$-band magnitude limit of $M_{\mathrm{r}} < -20$.
In all cases the inclusion of dust attenuation does not change the overall shape of the colour distributions, but leads to shifts to redder colours, of $\sim 0.18$ in the optical.
Astrid and Simba show similar colour distributions at all wavelengths, with a single strong peak in all distributions that corresponds to a predominantly blue star forming population.
Conversely, Swift-EAGLE and IllustrisTNG show more bimodal distributions, with a more pronounced red population at all wavelengths.

\subsubsection{Redshift evolution}

We have also generated rest-frame luminosity functions at a range of redshifts, shown in \fig{CV_lfs}, showing the evolution in abundance at different wavelengths\footnote{We stress that these rest-frame luminosity functions at higher redshift cannot be directly compared to observations without accounting for the necessary $K$-corrections.}.
In all galaxy formation models the FUV LF rises quickly, peaking at $z \sim2$, before falling again by $z = 0$.
This corresponds to the evolution in the cosmic star formation rate density \citep{madau_cosmic_2014}; more recent star formation leads to higher UV emission.
Interestingly, the shape of the LF also evolves with redshift in Simba, showing a more pronounced bright population at intermediate redshifts, before settling to a Schechter-like distribution by $z = 0$.
At redder wavelengths this fall in the LF after the peak of cosmic star formation is less pronounced, and in the $K$ band the evolution mirrors that of the galaxy stellar mass function, as expected.

\begin{figure*}
	\includegraphics[width=\textwidth]{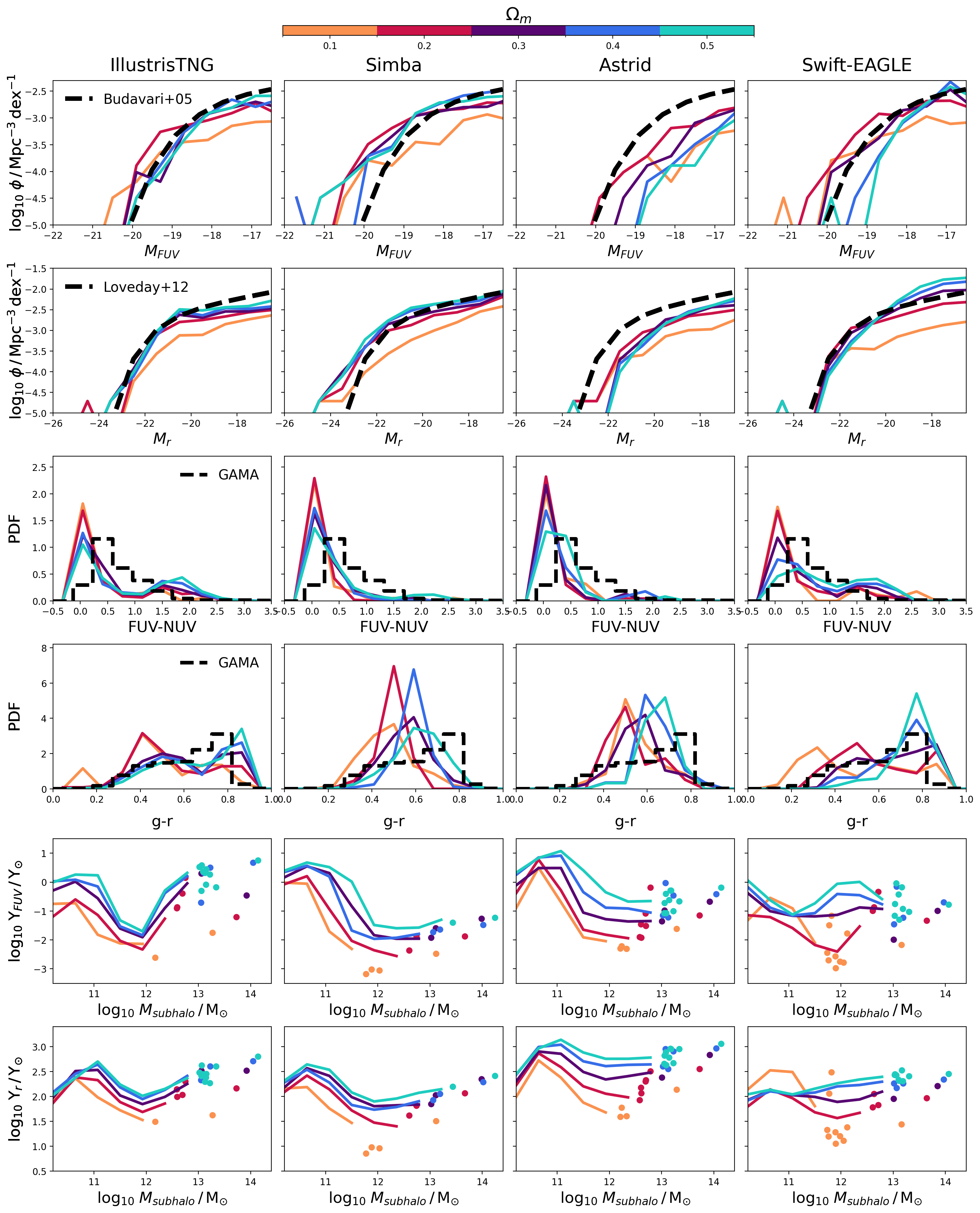}
    \caption{1P set variations of the dust-attenuated photometry at $z = 0.1$ for IllustrisTNG, Simba, Astrid and Swift-EAGLE (columns left to right, respectively), when changing \omegam.
    \textit{Top row:} GALEX FUV luminosity function, with observations from GALEX in the FUV \protect\citep{budavari_ultraviolet_2005}.
    \textit{Second row:} SDSS $r$-band luminosity function, with observations from the GAMA survey \protect\citep{loveday_galaxy_2012}.
    \textit{Third row:} GALEX FUV-NUV colour distribution, with observational constraints from GAMA.
    \textit{Fourth row:} SDSS $g-r$ colour distribution, with observational constraints from GAMA.
    \textit{Fifth row:} GALEX FUV subhalo mass-to-light ratio against halo mass (individual objects are plotted where there are fewer than 10 sources in a halo mass bin).
    \textit{Sixth row:} binned SDSS $r$-band subhalo mass-to-light ratio against halo mass.
    See \protect\sec{difficulties} for caveats on the observational constraints.}
    \label{fig:1P_omegam}
\end{figure*}

\subsection{1P set distributions}

We now explore the impact of parameter variations on LFs, colours and mass-to-light ratios by studying the simulations in the 1P set, which vary each parameter one by one around the fiducial parameters.

Figures \ref{fig:1P_omegam}, \ref{fig:1P_sigmae}, \ref{fig:1P_asnone}, \ref{fig:1P_asntwo}, \ref{fig:1P_aagnone} and \ref{fig:1P_aagntwo} show the GALEX $FUV$ and SDSS $r$-band LFs, the UV ($FUV - NUV$) and optical ($g - r$) colour distributions, and the FUV and $r$-band (subhalo) mass-to-light ratios from the 1P set for each galaxy formation model at $z = 0.1$.
We construct each mass-to-light (ML) ratio by normalising by the solar value at that wavelength,
\begin{align}
    \mathrm{log_{10}} \; \Upsilon_X \,/\, \Upsilon_{\odot} = \mathrm{log_{10}} \; (M_{\rm subhalo} \,/\, \mathrm{M_{\odot}}) \,/\, (M^{AB}_X \,/\, M^{AB}_{X,{\odot}}) \;,
\end{align}
where $M_{\rm subhalo}$ is the total subhalo mass, $M^{AB}_X$ is the AB absolute magnitude in band $X$, and $M^{AB}_{X,{\odot}}$ is the absolute magnitude of the sun in band $X$ \citep[values obtained from][]{willmer_absolute_2018}.

In order to provide a baseline for comparison we also show observational constraints from the GAMA survey \citep{loveday_galaxy_2012} in the $r$-band, GALEX in the FUV \citep{budavari_ultraviolet_2005}, and the UV ($FUV-NUV$) and optical ($g-r$) rest-frame colours from GAMA \citep{liske_galaxy_2015} as compiled by \cite{pandya_nature_2017}, all at $z \sim 0.1$.
We stress that these observationally derived rest--frame relations are not directly comparable to the results from CAMELS presented here, for reasons described in depth in \sec{difficulties}; here we just mention that these observations are necessarily $K$-corrected, and cover different volumes to those in CAMELS, making direct comparisons difficult.

From these figures it is clear that each parameter has a different impact depending on the wavelength and galaxy formation model considered.
We first explore the cosmological parameters, before moving on to the impact of the astrophysical parameters.

\subsubsection{Changes in \omegam}
\label{sec:1P_omegam}

An increase in \omegam\ leads to an increase in the overall normalisation of the halo mass function \citep{villaescusa-navarro_camels_2021}, due to accelerated formation of more massive haloes \citep{mcclintock_aemulus_2019}.
However, it has been unclear how this translates into changes in the LFs beyond the increase in the halo number density, since the emission is directly dependent on the star formation and metal enrichment history of each galaxy, which itself depends in a complicated way on the underlying cosmology \citep[][, \textit{in prep.}]{iyer_diversity_2020}.

We see in \fig{1P_omegam} that increasing \omegam\ leads to an increase in the number of faint galaxies in the optical ($r$-band), with little dependence on the galaxy formation model, similar to how it impacts the galaxy stellar mass function \citep{ni_camels_2023,lovell_hierarchy_2023,jo_calibrating_2023}.
We see a similar relationship in the UV, though the relations are more noisy, which reflects the sensitivity of the UV emission to star formation on relatively short timescales \citep[last $\sim100 \; \mathrm{Myr}$;][]{lee_comparison_2009}.
In general, however, the optical and UV LFs are degenerate where \omegam\ > 0.3, with very little differences across all magnitudes.
The UV and optical ML ratios in \fig{1P_omegam} help explain this behaviour.
Whilst the overall normalisation of the halo mass function increases with increasing \omegam, the ML ratio also increases; galaxies in equal mass haloes are fainter in a universe with a higher \omegam.
This trend is monotonic for all models, except Swift-EAGLE, which shows a more complicated evolution for \omegam = 0.1 (the ML ratio in the UV and optical is much higher in low mass haloes).

Above a halo mass of $10^{11} \; \mathrm{M_{\odot}}$ there is a general trend across all simulations and parameters for the ML ratio to decrease with increasing subhalo mass; this explains the characteristic schechter-like shape of the luminosity function in these bands at faint magnitudes \citep{bullock_small-scale_2017}.
The mass at which this fall is somewhat sensitive to \omegam, which explains the correlation of the shape of the LFs with \omegam\ in all models.
The magnitude of the dependence of the ML ratio on \omegam\ is also model dependent; the largest variations at a halo mass of $10^{12}\;\mathrm{M_\odot}$ are seen in Swift-EAGLE and Astrid, with a difference in the ML ratio in the UV of $|\Upsilon_X | > 2$.
This explains why there is also a significant decrease in the normalisation at the \textit{bright} end of the UV LF for the highest \omegam\ values in these models.

The colour distributions also tell an interesting story.
As \omegam\ increases, we see redder colour distributions in both the UV and optical.
The magnitude of this effect in the optical is largest for the Swift-EAGLE model; the blue population centred at $g-r = 0.3$ for \omegam\ = 0.1 completely disappears when \omegam\ is increased to 0.5, leading to a unimodal distribution of red galaxies ($g-r \sim 0.8$).
This behaviour suggests that not only is the abundance of haloes and their ML ratios changing with \omegam, but also that this effect is wavelength dependent.
Also interesting is that there are less degeneracies between models with high \omegam\ in the UV and optical colour spaces, with clear differences in their colour distributions throughout the range of \omegam\ probed.
This will become important when performing SBI to infer the value of \omegam\ from these distributions, discussed further in \sec{feature_importance}.

Since a higher \omegam\ leads to not only a higher abundance of haloes but also their accelerated formation \citep{mcclintock_aemulus_2019}, the stellar ages in these massive haloes may be generally higher, leading to the reduced abundance of UV-bright galaxies, and an overall reddening of their colours.
Quiescent fractions have not been studied in the fiducial Astrid simulation at $z = 0.1$, but in EAGLE they show a strong correlation with galaxy mass \citep{furlong_evolution_2015}, in agreement with observational constraints on the passive fraction, and supporting this narrative.
However, similar behaviour is also seen in IllustrisTNG \citep{donnari_star_2019,donnari_quenched_2021,gabrielpillai_galaxy_2022} and Simba \citep{dave_simba:_2019,rodriguez_montero_mergers_2019,appleby_impact_2020}.
Simba produces smoother star formation histories that tend to form later \citep{dickey_iq_2021}, which may also explain the elevated recent SFRs at higher \omegam, since any long-term variability will have had less time to imprint on the overall SFH \citep{zheng_rapidly_2022}.

Mergers between haloes and their host galaxies are also more frequent in a universe with higher \omegam.
\cite{rodriguez_montero_mergers_2019} showed how mergers have very little effect on passive fractions in Simba, suggesting this effect will have little impact on recent star formation, and therefore the UV emission, in this model.
\cite{dickey_iq_2021} showed how passive fractions at low-masses are elevated in isolated galaxies in EAGLE compared to IllustrisTNG and Simba, but lower for the most massive objects, which makes it difficult to pick apart the impact of (recent) mergers in these models.

\subsubsection{Changes in \sigmae}
\label{sec:1P_sigmae}

\begin{figure*}
	\includegraphics[width=\textwidth]{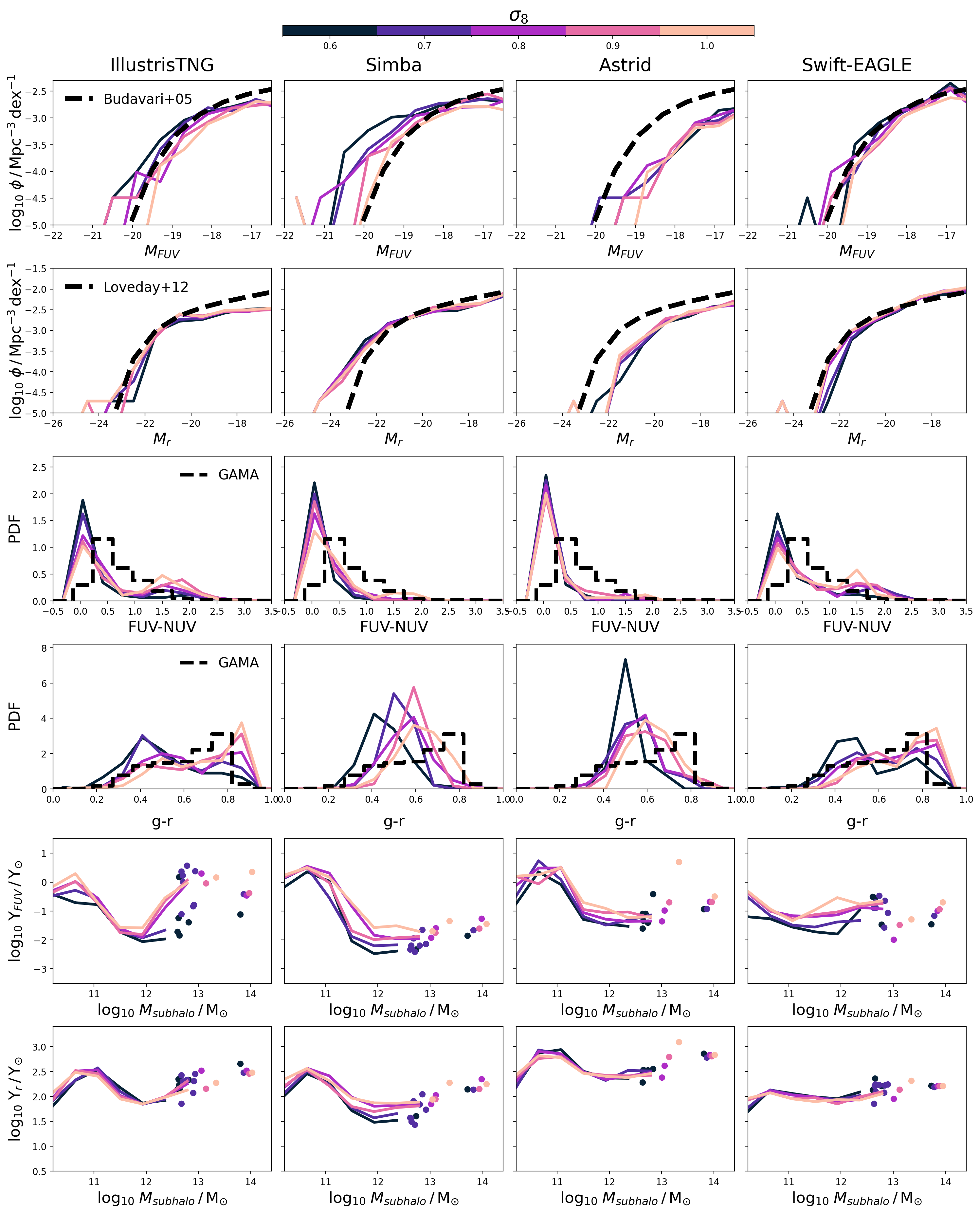}
    \caption{
        The same as \fig{1P_omegam}, but showing the variation in the \sigmae\ 1P set.
    }
    \label{fig:1P_sigmae}
\end{figure*}

The value of \sigmae\ encodes the degree of matter clustering, and it has an interesting impact on the LFs and colours.
Increasing \sigmae\ does not have a strong effect on the optical LFs in all models, nor the optical ML ratios (except a small increase for increasing \omegam\ in Simba at halo masses $\sim 10^{12} \; \mathrm{M_\odot}$).
This is somewhat surprising, given that \sigmae\ has a known impact on the high mass end of the halo mass function \citep[hence the utility of cluster counts for constraining \sigmae;][]{press_formation_1974,tinker_toward_2008}.
However, \sigmae\ does have a discernible effect on the UV LF in some models.
This effect is most pronounced in Simba, where the magnitude of the brightest object in the UV increases by up to 1.5 magnitudes to $M_{\rm FUV} = -20$ for \sigmae\ = 0.6. 
This can be explained by increases in the UV ML ratios of up to 0.8 dex in Simba as \sigmae\ increases, much greater than any other model.

However, the strongest dependence on \sigmae\ is seen in the colours; increasing \sigmae\ tends to lead to redder colours in the UV, but particularly in the optical ($g-r$), for all models.
In Swift-EAGLE and IllustrisTNG this manifests as a relative shift in the height of the red and blue populations in their respective bimodal colour distributions.
In contrast, in Simba and Astrid this manifests as a shift in the peak of the unimodal distribution. 
This suggests that there is a fundamental difference in how changes in \sigmae\ manifest in differences in the colour distributions.

\begin{figure*}
    \includegraphics[width=\textwidth]{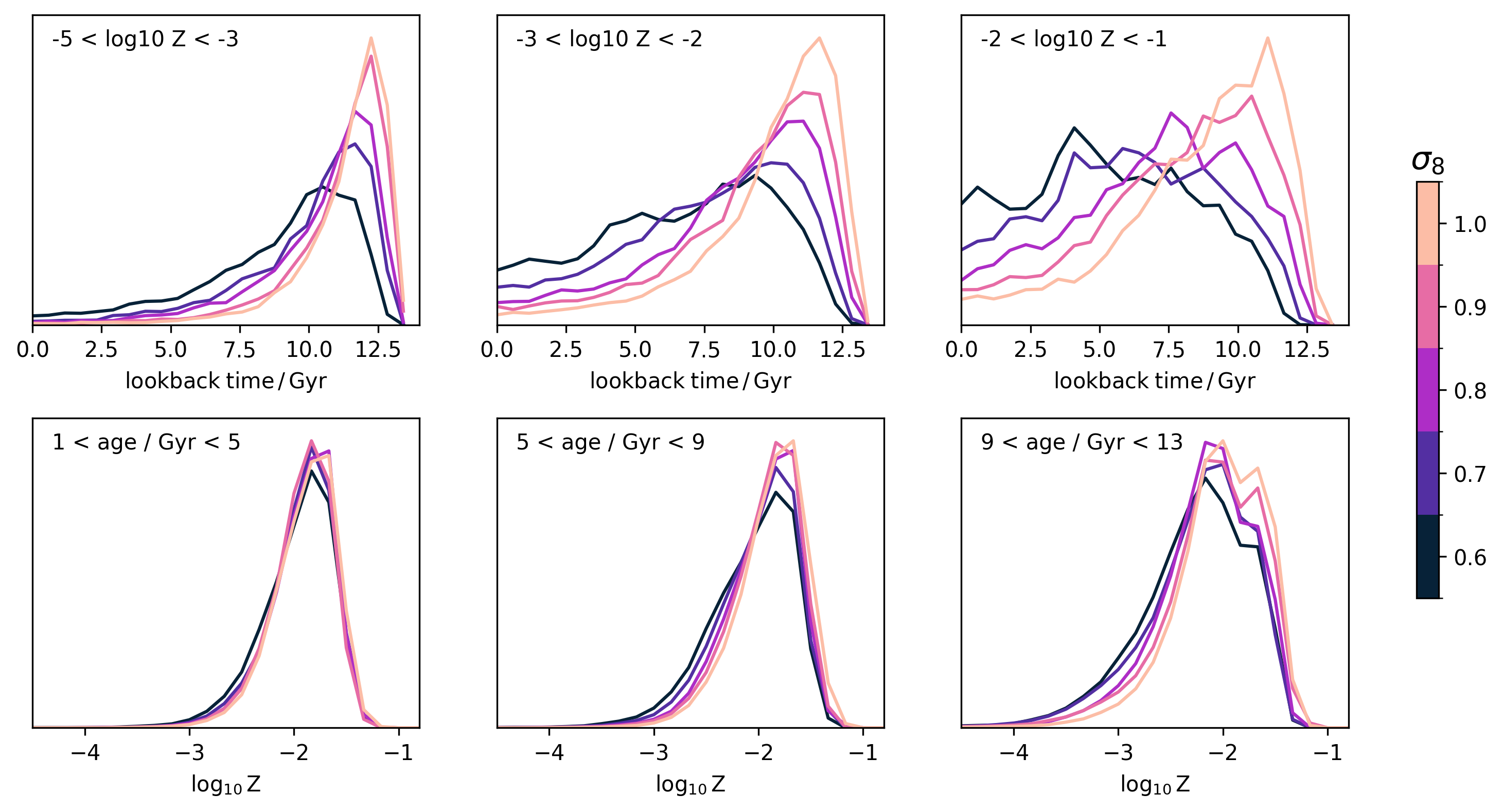}
\caption{\textit{Top:} normalised cosmic star formation history in bins of stellar metallicity, from the \textsc{IllustrisTNG} 1P set, changing \sigmae. \textit{Bottom:} normalised stellar metallicity distribution in bins of stellar age.}
\label{fig:sfzh}
\end{figure*}

Since the emission is wholly dependent on the star formation and metal enrichment history (SFZH) of our galaxies, we explore how this SFZH, and its marginal distributions, is dependent on the value of \sigmae\ to try and understand these correlations. 
\fig{sfzh} shows the normalised star formation history, using the summed ages of all star particles in each volume, from the IllustrisTNG 1P set changing \sigmae, further binned by stellar metallicity.
It is clear that, for higher \sigmae, galaxies form earlier, and the aggregate stellar population age is therefore older.
We find a similar trend for the other models.
This is the primary driver of the redder colours seen in \fig{1P_sigmae}.

\fig{sfzh} also shows the normalised metallicity distributions as a function of \sigmae\ in IllustrisTNG, in bins of stellar age.
These show a more subtle dependence on \sigmae; greater clustering leads to marginally higher metallicites, but only in the oldest stellar populations.
This may have a second order impact on the colour distributions, but the main driver is the stellar ages.
A more detailed exploration of the impact of cosmological and astrophysical parameters on the star formation and metal enrichment histories of galaxies in CAMELS is provided in Iyer et al. \textit{in prep.}.

For larger \sigmae, galaxies of all masses will tend to live in denser environments, closer to massive haloes and filaments.
This will impact the accretion rate of subhaloes, as well as the accretion rate of satellite subhaloes into larger host haloes.
These processes may have an additional impact on galaxy colours.
\cite{bulichi_how_2024} explored how this impacts galaxy properties in the Simba model; they found that SFRs tend to be reduced with increasing proximity to massive neighbours, that satellites are affected more than centrals, and that increased shock heating in large scale structures such as filaments, rather than the impact of feedback, drives these reductions in SFR.
\cite{bulichi_how_2024} also compared their results for Simba with EAGLE and Illustris, and found that the trend of SFR with filament proximity is smoother in Simba, compared to a more abrupt change in EAGLE and IllustrisTNG.
These effects may explain the bi-modal colour distributions seen in these latter models, and contribute to the relative impact of changes in \sigmae\ on the colour distributions.

\subsubsection{Changes in supernovae feedback parameters}
\label{sec:1P_sn}

\begin{figure*}
	\includegraphics[width=\textwidth]{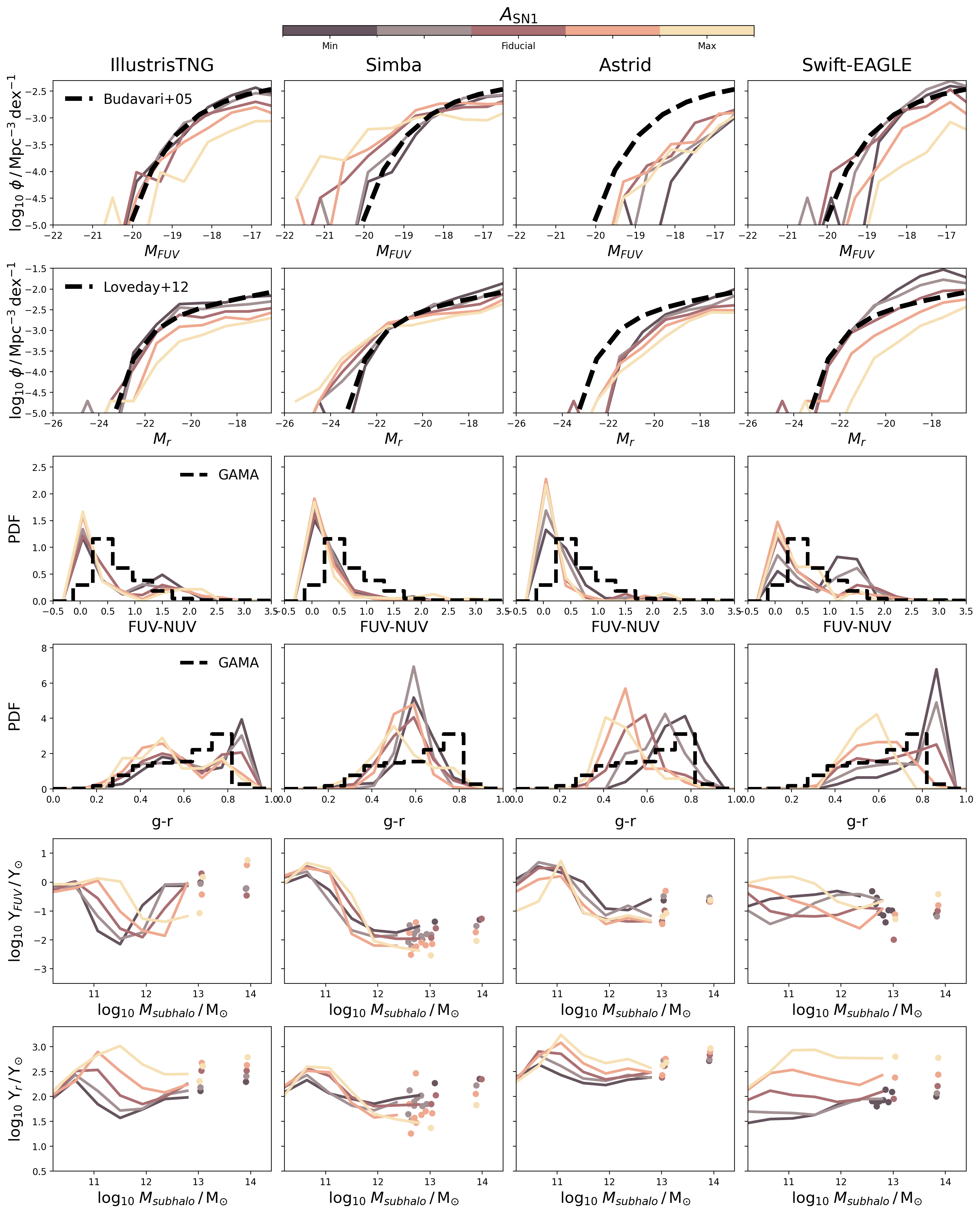}
    \caption{
        The same as \fig{1P_omegam}, but showing the variation in the \asnone\ 1P set.
    }
    \label{fig:1P_asnone}
\end{figure*}

\begin{figure*}
	\includegraphics[width=\textwidth]{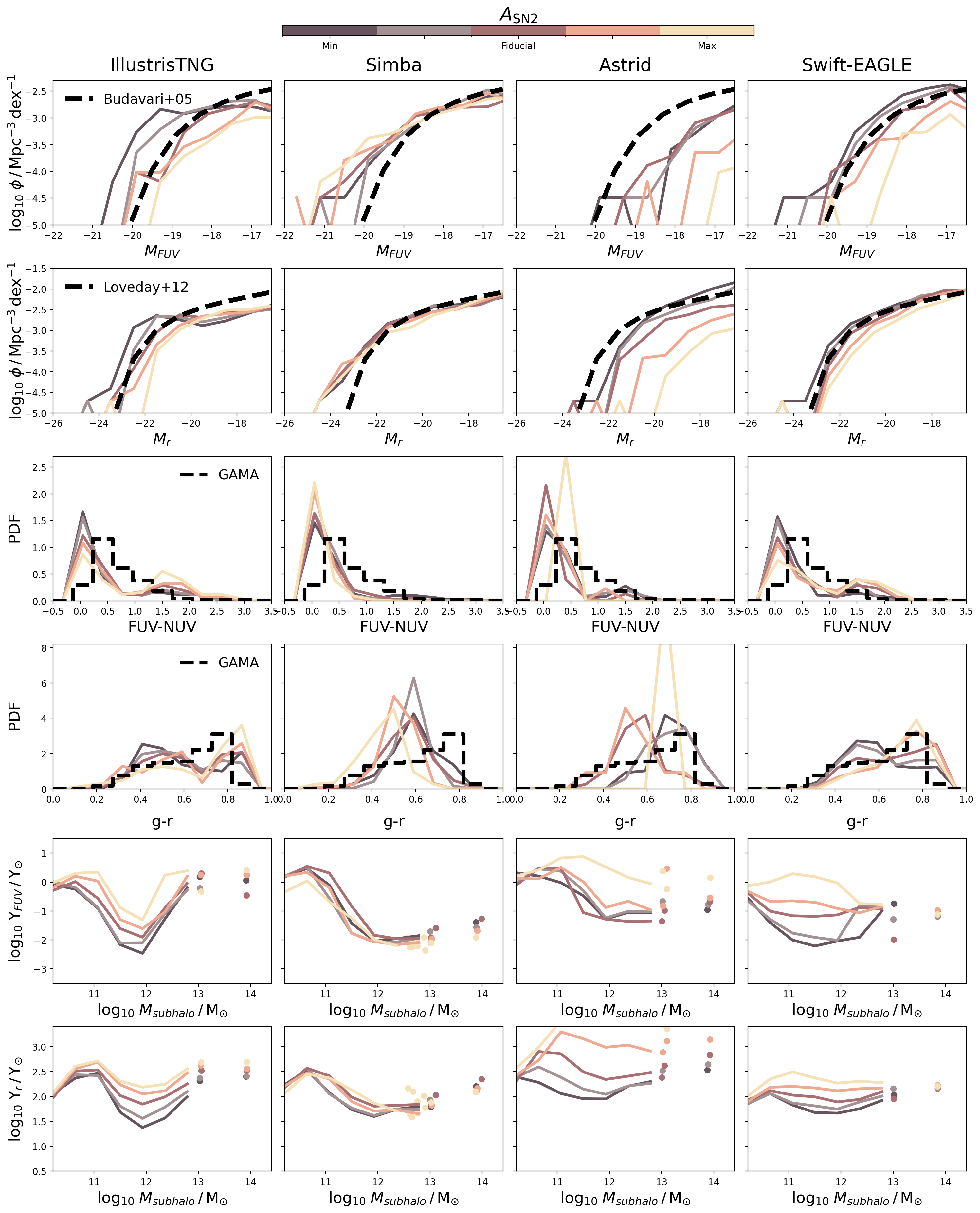}
    \caption{
        The same as \fig{1P_omegam}, but showing the variation in the \asntwo\ 1P set.
    }
    \label{fig:1P_asntwo}
\end{figure*}

The \asnone\ and \asntwo\ parameters control different elements of the subgrid star formation feedback models in each galaxy formation model (see \tab{parameters} for a summary), and as such it is not possible to compare them directly.
Instead we explore each galaxy formation model in turn, and discuss the impact of each parameter change on that specific model.
\fig{1P_asnone} show the variation of the LFs, colour distributions and ML ratios for \asnone, and \fig{1P_asntwo} shows the same for \asntwo.

In IllustrisTNG, \asnone\ controls the feedback energy per unit SFR in the form of stellar winds, whereas \asntwo\ controls the speed of those winds.
Both parameters have a significant impact on the UV and optical LFs.
In the optical, increasing \asnone\ tends to lead to a reduction in the normalisation at all magnitudes, which is reflected in the increasing normalisation of the ML ratio at all halo masses with increasing \asnone.
This increase is greatest for haloes with mass between $10^{11} - 10^{12} \, \mathrm{M_{\odot}}$, which reflects the fact that feedback from stellar winds affects these relatively lower mass haloes more effectively.
In the UV, the evolution is more complex; the UV ML ratio increases with increasing \asnone\ at low halo masses ($< 10^{12} \, \mathrm{M_{\odot}}$), but above this mass limit the relation reverses.
This leads to a reduction in the normalisation of the faint end of the UV LF, but an increase in the normalisation at the bright end.

This behaviour may be explained by the secondary effect of stellar feedback suppressing supermassive black hole (SMBH) growth; \cite{tillman_exploration_2023} show how increasing \asnone\ leads to a decrease in the overall number density of black holes, reducing the global impact of AGN feedback.
This latter effect also has an obvious impact on the colours, leading to a much bluer distribution in the UV-optical.
It may also reflect that increasing \asnone\ leads to greater gas heating, rather than expelling gas from haloes, so that gas is still available for star formation at later times once it has cooled.

On the contrary, \asntwo, which controls the speed of galactic winds, has a large impact on the bright end of each LF.
This is reflected in the ML ratios, where there is a strong dependence on \asntwo\ in relatively higher mass haloes ($> 10^{12} \, \mathrm{M_{\odot}}$)
This may reflect the fact that gas has been fully expelled from haloes by the higher wind speeds, reducing the overall gas fractions (though the details of baryon spread can depend in a complex way on feedback parameters, see \citealt{gebhardt_cosmological_2024}).
Impacts on the growth and accretion onto nearby massive haloes, containing the brightest galaxies, may also be a secondary effect.
For \asntwo\ the secondary effects on the black hole number density are far less prominent.

In Simba, \asnone\ controls the mass loading of stellar winds, and \asntwo\ controls the wind speed.
However, contrary to IllustrisTNG, increasing the wind speed (\asntwo) leads to \textit{brighter} galaxies in the UV for higher wind speeds.
The ML ratios reflect this, showing a turnover in the \asnone\ dependence above $10^{12} \, \mathrm{M_{\odot}}$.
\cite{gebhardt_cosmological_2024} showed how increased wind speed actually leads to reduced baryon spread, since less gas is available in the central regions for AGN accretion, which may explain why there is increased star formation leading to higher UV emission.
\cite{tillman_exploration_2023} showed that \asntwo\ in Simba also curtails SMBH growth, which would contribute to the bright end behaviour of the LFs, and also the bluer optical colour distributions.
Also counter intuitively, increasing the mass loading of winds (\asnone) leads to an \textit{increase} in the number of UV bright galaxies, and a decrease in the low-mass normalisation, which is also reflected in the increased fraction of blue galaxies.

In Astrid, \asnone\ controls the feedback energy per unit SFR in stellar winds, and \asntwo\ the wind speed, both similarly to IllustrisTNG.
\asnone\ slightly reduces the faint-end abundance in the optical, due to increased ML ratios for higher \asnone\ below $10^{12} \, \mathrm{M_{\odot}}$.
In the UV it is difficult to discern any clear trends due to the low number of galaxies, however the UV ML ratios show the opposite trend to the optical above $10^{12} \, \mathrm{M_{\odot}}$, with higher ML ratios for lower \asnone.
asntwo\ has a very strong impact on the global normalisation of the LF in the UV and optical, leading to a drop of over 1 dex across all magnitudes.
The implementation of the wind speed affects both high and low mass haloes, as clearly seen in the ML ratios, reducing star formation and the relative abundance across all halo masses.
As a result, the number of galaxies in Astrid that reach the stellar mass and $r$-band magnitude brightness limits for the colour plots in \fig{1P_asntwo} is significantly reduced for high \asntwo, leading to very noisy distributions; despite this the overall trends are for bluer distributions at higher feedback parameters.

Finally, in Swift-EAGLE \asnone\ controls the energy per unit SNII feedback event, which drive galactic winds, and \asntwo\ controls the metallicity dependence of the feedback fraction per unit mass, $n_Z$, 
\begin{align}
    f_{\rm th} = f_{\rm th,min} + \dfrac{f_{\rm th,max} - f_{\rm th,min}}{1 + \left( \dfrac{Z}{0.1 \, Z_{\odot}} \right)^{n_Z} \, \left( \dfrac{n_{\rm H, birth}}{n_{\rm H,0}} \right)^{-n_n} } \,\,,
\end{align}
where $f_{\rm th,max}$ and $f_{\rm th,min}$ are the upper and lower limits of the feedback fraction, and the $(n_{\mathrm{H,birth}} / n_{\mathrm{H,0}})^{-n_n}$ term controls the density dependence of the feedback fraction.
Increasing \asnone\ leads to an overall reduction in the UV and optical LF normalisation.
In the optical this is clearly explained by the monotonic relationship between ML ratio and \asnone, however in the UV things are more complicated, with clear dependencies of UV ML on halo mass that further depend on the value of \asnone.
As for the colours, increasing \asnone\ leads to bluer galaxy distributions; this may reflect secondary effects on the growth of SMBHs, as discussed previously, or the fact that gas is heated rather than expelled, meaning there is fuel for later star formation. 
Increasing \asntwo\ leads to increased ML ratios in the optical and UV, and subsequently reduced normalisation of the LFs at all magnitudes.
It also leads to redder distributions, which reflects the fact that galaxies that enrich early are then susceptible to more powerful feedback at later times, reducing their star formation.

\subsubsection{Changes in AGN feedback parameters}
\label{sec:1P_agn}

\begin{figure*}
	\includegraphics[width=\textwidth]{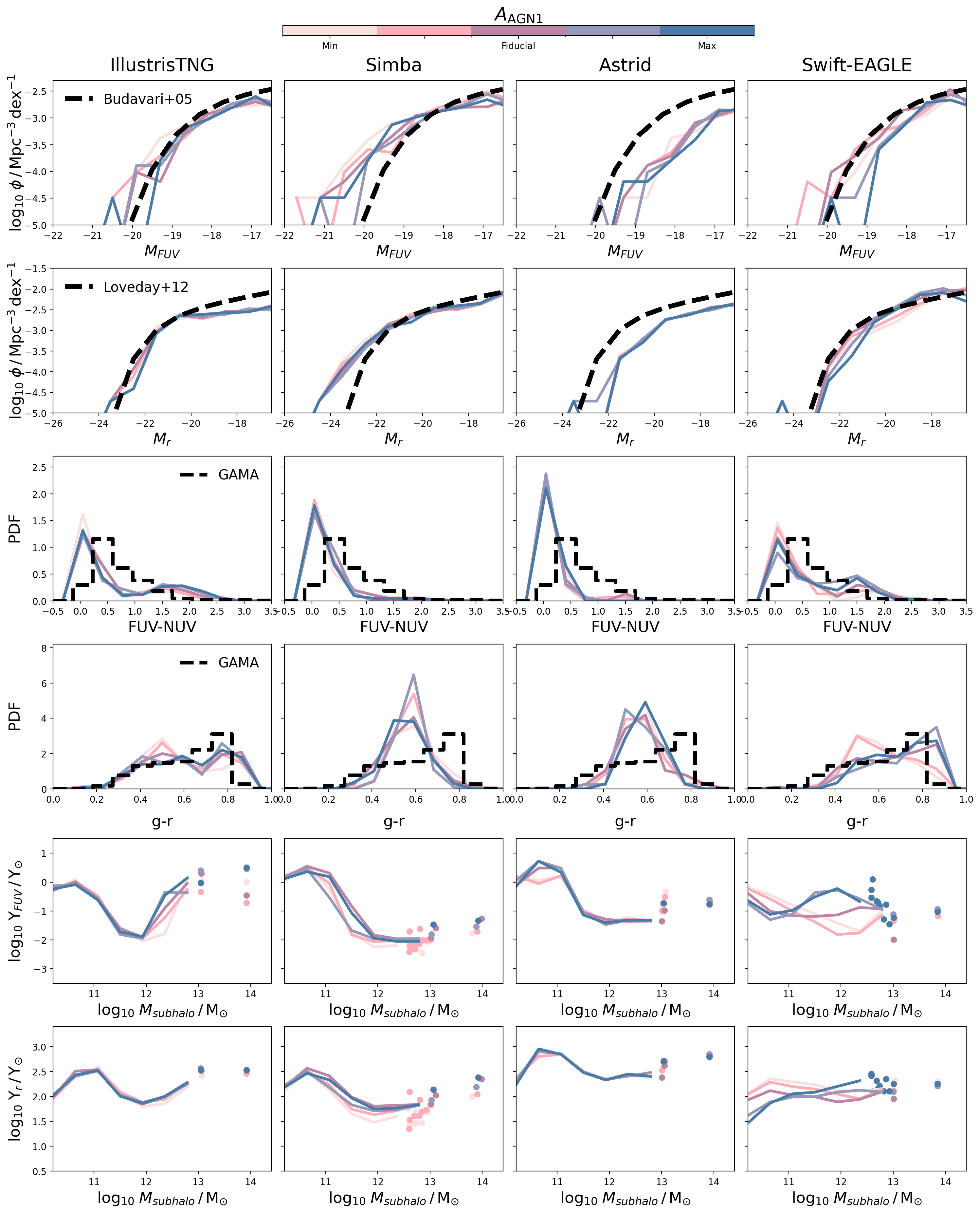}
    \caption{
        The same as \fig{1P_omegam}, but showing the variation in the \aagnone\ 1P set.
    }
    \label{fig:1P_aagnone}
\end{figure*}

\begin{figure*}
	\includegraphics[width=\textwidth]{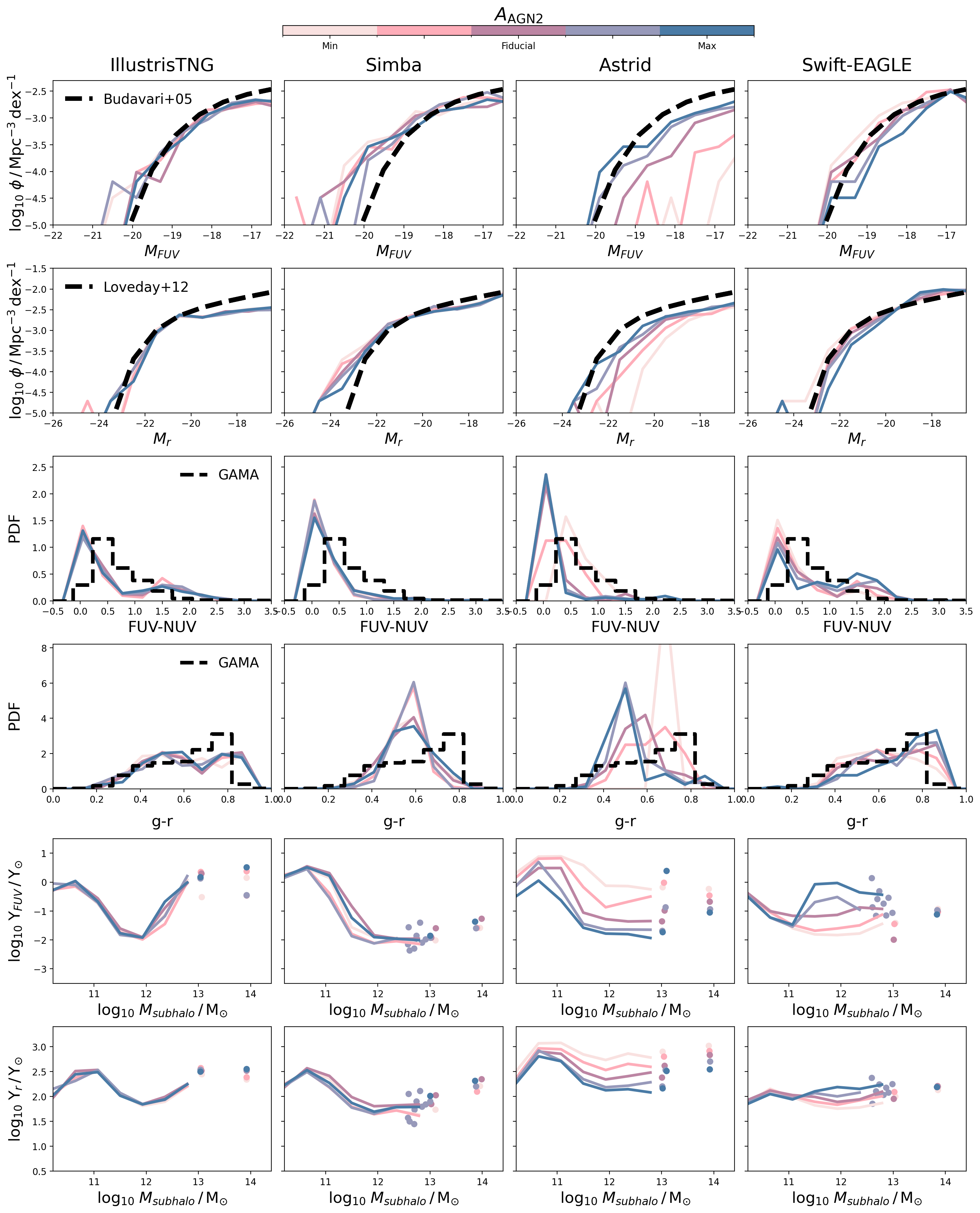}
    \caption{
        The same as \fig{1P_omegam}, but showing the variation in the \aagntwo\ 1P set.
    }
    \label{fig:1P_aagntwo}
\end{figure*}

As is the case for the supernovae feedback parameters, \aagnone\ and \aagntwo\ control different elements of the subgrid AGN feedback models in each galaxy formation model (see \tab{parameters} for a summary), and cannot be compared directly.
Contrary to the supernovae parameters, the AGN parameters have a much reduced effect in almost all the galaxy formation models.
This reflects the small box size in CAMELS ($(25 \; h^{-1})^3 \; \mathrm{Mpc^3}$) and subsequent lack of (relatively) massive haloes, as well as SMBH self-regulation.

In IllustrisTNG, \aagnone\ controls the kinetic energy released per unit SMBH accretion mass, and \aagntwo\ the ejection speed and burstiness of that ejected material (where greater burstiness leads to more frequent but lower energy feedback events; \citealt{tillman_exploration_2023}).
However, neither has any appreciable effect in either the LFs or colour distributions above and beyond butterfly effects \citep{genel_quantification_2019}.
A similar story unfolds for Simba; in this model \aagnone\ controls the momentum in quasar and jet mode feedback, and \aagntwo\ the speed of the jet.
However, neither has any impact on the LFs and colours.

In Astrid \aagnone\ controls energy per unit SMBH accretion, and this particular parameter also has no impact on the LFs and colours.
However, \aagntwo, which controls the energy injected per unit accretion mass in the \textit{thermal} feedback mode, has a large impact on the LFs, leading to an increase in excess of 1.5 dex at the bright end of the UV LF, and 1 dex in the optical.
Why does increasing this particular channel of BH feedback lead to \textit{increased} number densities?
The reason is efficient self-regulation: enhanced thermal feedback severely suppresses the growth of SMBHs, reducing the overall feedback injected (particularly in the jet mode), leading to enhanced star formation \citep{ni_camels_2023}.
This also leads to bluer colours in both the UV and optical.
Another reason why we see a stronger effect from the AGN parameters in Astrid compared to Simba and IllustrisTNG is that Astrid has a higher number density of massive black holes in the fiducial model, as evidenced by the 0.5 dex higher normalisation of the black hole mass function at the high-mass end \citep{ni_camels_2023}.

Finally, in Swift-EAGLE AGN feedback is implemented in a purely thermal mode; \aagnone\ controls the efficiency scaling of the Bondi accretion rate, and \aagntwo\ the temperature jump of the gas during AGN feedback events.
Neither parameter has a large effect on the optical LFs, but there is a small difference in the bright end in the UV, whereby increasing both parameters reduces the abundance of UV bright galaxies.
The differences in the colours are more pronounced; increasing both parameters lead to redder UV and optical colours.
This suggests that increasing the primary efficacy of thermal feedback in Swift-EAGLE does not lead to the same self-regulation effects seen in Astrid, and had a more predictable effect on the luminosities and colours of massive, bright galaxies hosting AGN.

\section{Simulation based inference with LtU-ILI}
\label{sec:sbi}

We now explore the use of our forward modelled UV-optical luminosity and colour distributions from the CAMELS LH set within an SBI framework.
SBI is a powerful framework for performing inference in this domain since it permits amortized posteriors, which significantly speeds up the process of obtaining posteriors.
This allows us to apply the trained model to hundreds of test set simulations in just fractions of a second.
Additionally, SBI does not require an explicit likelihood, which permit us to use more complex noise models, which we will explore in future work when comparing directly to observations.

\subsection{SBI Architecture}
\label{sec:architecture}

We use the LtU-ILI package \citep{ho_ltu-ili_2024} to perform SBI. 
LtU-ILI is a codebase that simplifies many of the steps in an SBI analysis, particularly those concerning testing and validation such as estimating the epistemic uncertainty, choosing hyperparameters, and assessing model misspecification.
We use the \textsc{sbi} package within LtU-ILI, and perform neural posterior estimation \citep[NPE;][]{papamakarios_fast_2016,greenberg_automatic_2019}, to allow amortized inference.
We construct an ensemble of two models, each composed of a neural spline flow \citep{durkan_neural_2019} with 4 transforms, 10 spline bins, and 60 hidden features.
During training we use a batch size of 4, and a fixed learning rate of $5 \times 10^{-4}$, and an improvement based stopping criterion. 

Throughout the rest of this manuscript we will present models trained and applied to different simulations.
We use the same architecture for each trained model.
For each simulation we reserve a subset of the LH simulations as a test set (10\%), and a further subset of the remaining simulations for validation (10\%). 
The remaining simulations we use for training, and evaluate our model performance during training on the validation set.
Our priors are set by the chosen distribution of the parameters in the LH set, described in \cite{villaescusa-navarro_camels_2021}.
After training, to produce posteriors for a given set of parameters we directly take 1000 samples from the NPE.

\subsection{Feature Vector}
For our input feature vector we choose a subset of attenuated luminosity functions (in the GALEX $FUV$ and $NUV$ bands, and the SDSS $ugriz$ bands) and colour distributions ($FUV-NUV$, $g-r$, $r-i$, $i-z$).
For each luminosity function we apply 12 bins linearly spaced between bright- and faint--end AB magnitude limits given in \tab{lf_limits}.
For colour distributions, we also apply 12 bins linearly spaced between the red- and blue--limits given in \tab{colour_limits}.
We use the same cuts in mass and magnitude as used in the training data when performing inference.
We provide the observational data as a single one-dimensional concatenated vector, $\mathbf{X}$; where multiple redshifts are considered these are simply concatenated together.
We perform coverage tests for each trained model, discussed in \app{coverage}.
In the following sections we will explore how our results depend on the choice of luminosity functions or colours, and how combining distributions at different redshifts can improve constraints on key cosmological and astrophysical parameters.

\begin{table}
	\centering
	\caption{Bright- and faint--end AB magnitude limits chosen for luminosity functions used in the SBI analysis. All luminosity functions are binned in 12 bins linearly spaced between these limits.}
	\label{tab:lf_limits}
	\begin{tabular}{lccr} 
		\hline
		Band & Bright limit & Faint limit \\
		\hline
        FUV & -20.5 & -15 \\
        NUV & -20.5 & -15 \\
		\textit{u} & -21.5 & -16 \\
		\textit{g} & -22.5 & -17 \\
		\textit{r} & -23.5 & -18 \\
        \textit{i} & -24 & -18.5 \\
        \textit{z} & -24.5 & -19 \\
		\hline
	\end{tabular}
\end{table}

\begin{table}
	\centering
	\caption{Blue- and red--limiting colours chosen for the construction of colour distributions used in the SBI analysis. All colour distributions are binned in 12 bins linearly spaced between these limits.}
	\label{tab:colour_limits}
	\begin{tabular}{lccr} 
		\hline
		Colour & Bright limit & Faint limit \\
		\hline
        $FUV$ -- $NUV$ & -0.5 & 3.5 \\
		$g$ -- $r$ & 0.0 & 1.0 \\
		$r$ -- $i$ & 0.0 & 0.5 \\
        $i$ -- $z$ & -0.1 & 0.4 \\
		\hline
	\end{tabular}
\end{table}





\subsection{Inference}
\label{sec:single_sim}

\begin{figure*}
	\includegraphics[width=0.95\textwidth]{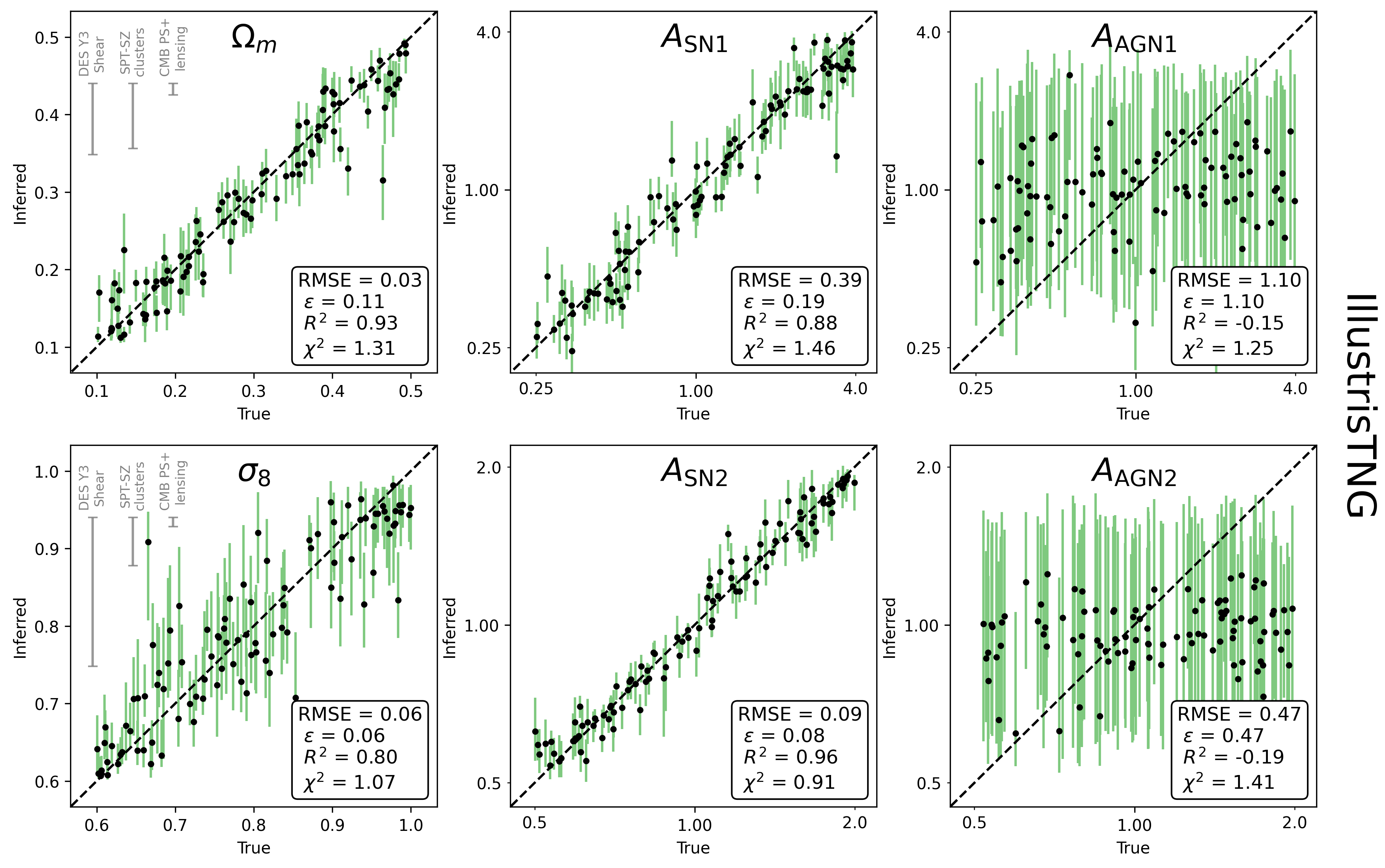}
    \includegraphics[width=0.95\textwidth]{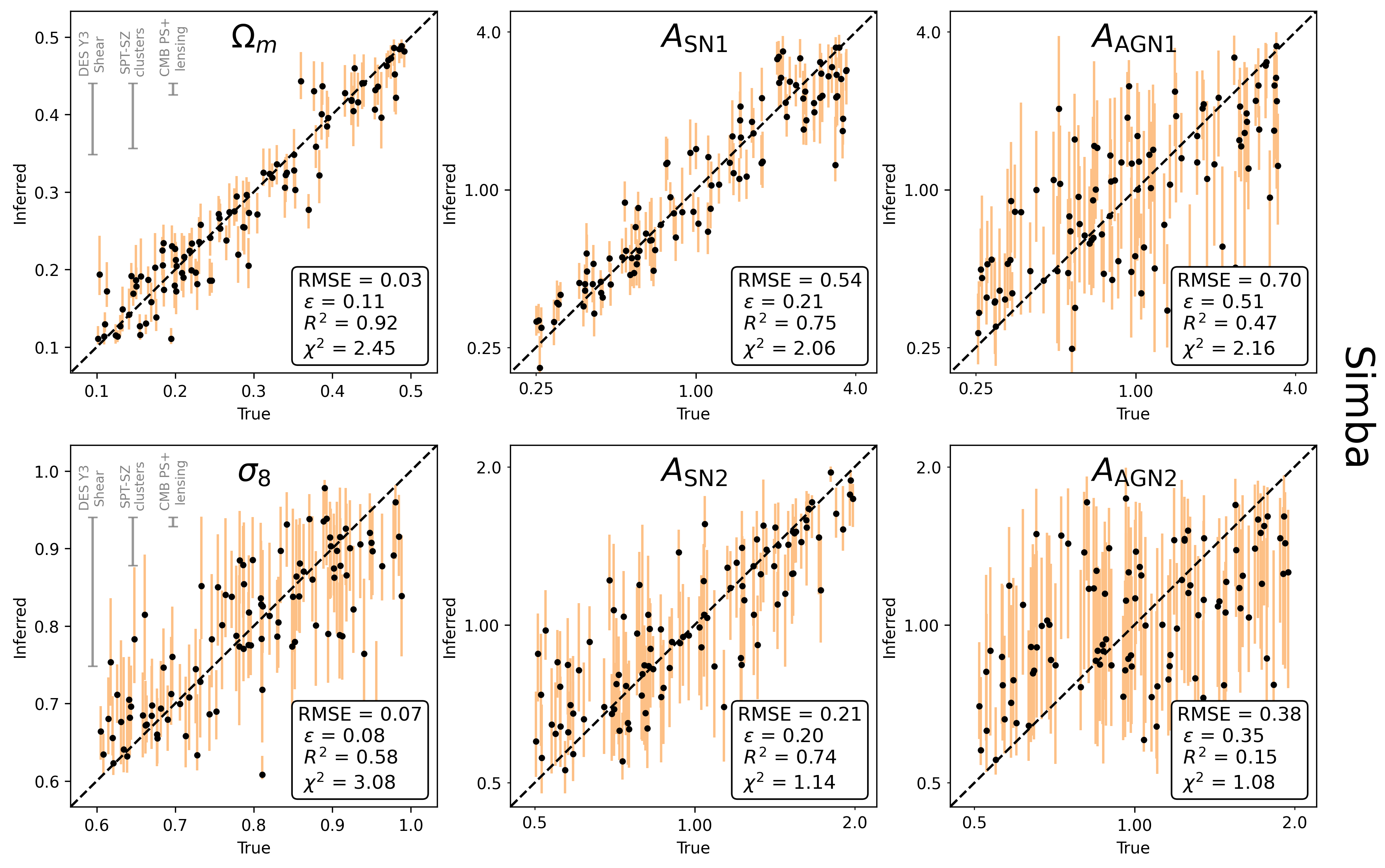}
    \caption{Predicted marginal posteriors on the test set vs the true values for each cosmological (\omegam, \sigmae) and astrophysical (\asnone, \asntwo, \aagnone, \aagntwo) parameter. These estimates use both colours and luminosity functions at $z = 0.1$, described in more detail in \sec{single_sim}, using training and testing data from the same galaxy formation model; we discuss generalisability across different models in \sec{robustness}.
    We also show the 16$^{\text{th}}$ -- 84$^{\text{th}}$ percentile range of the posterior constraints on \omegam\ and \sigmae\ from the Planck 2018 CMB analysis \citep[including the lensing reconstruction;][]{planck_collaboration_planck_2020}, the Dark Energy Survey (DES) Year 3 angular power spectra cosmic shear analysis \citep{doux_dark_2022}, and the South Pole Telescope (SPT) Sunyaev-Zel'dovich selected cluster abundance analysis \citep{de_haan_cosmological_2016}.
    \textit{Top two rows}: \textsc{IllustrisTNG}. \textit{Bottom two rows}: \textsc{Simba}.
    }
    \label{fig:single_predictions_A}
\end{figure*}

\begin{figure*}
    \includegraphics[width=0.95\textwidth]{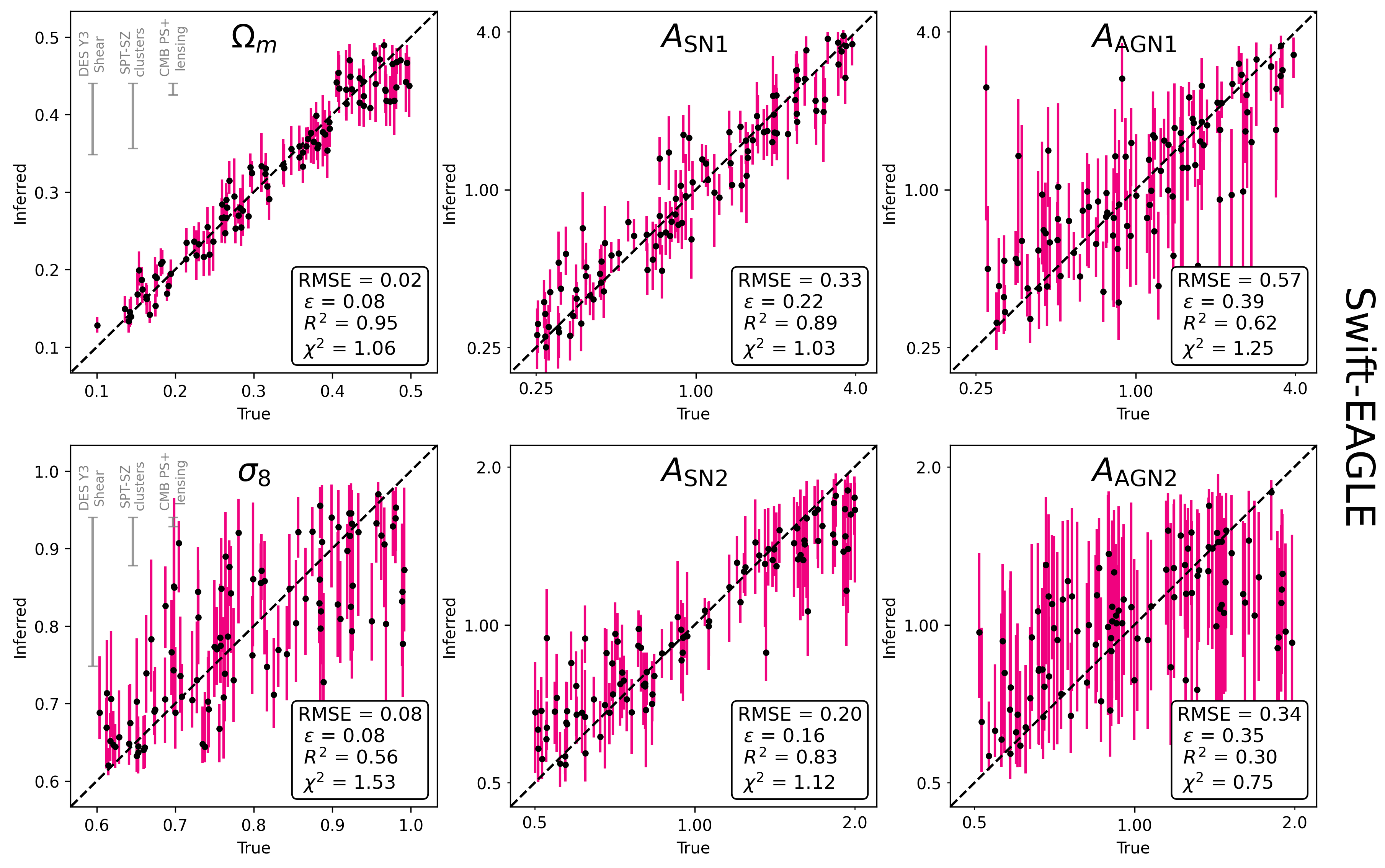}
    \includegraphics[width=0.95\textwidth]{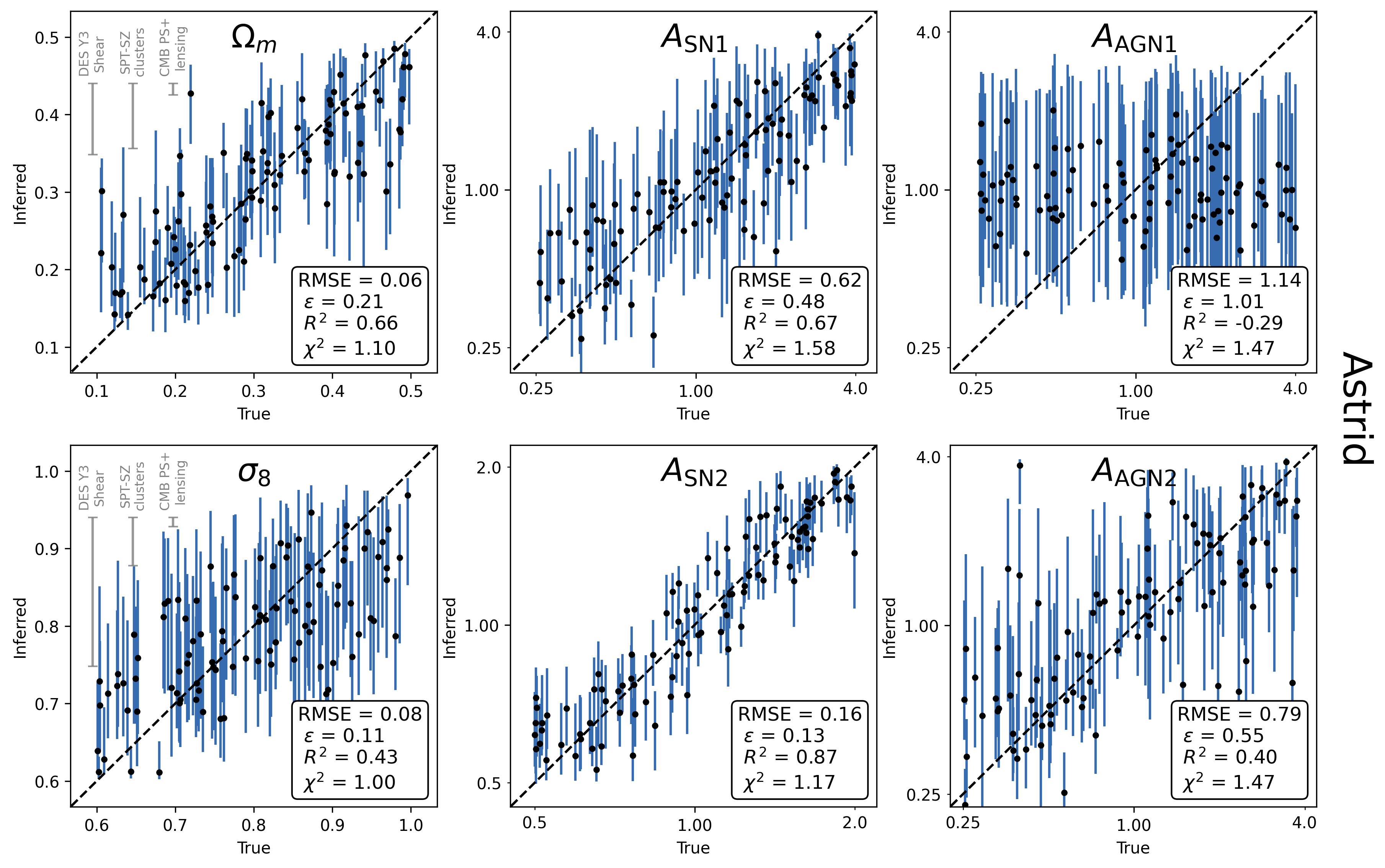}
    \caption{The same as \fig{single_predictions_A}, but showing predictions for \textsc{Swift-EAGLE} (top two rows) and \textsc{Astrid} (bottom two rows).}
    \label{fig:single_predictions_B}
\end{figure*}

We begin by showing the predicted marginal posteriors on our cosmological and astrophysical parameters for all test set simulations, using a model trained and tested on the same galaxy formation model, in Figures \ref{fig:single_predictions_A} \& \ref{fig:single_predictions_B}.
We show posterior predictions for \omegam, \sigmae, \asnone, \asntwo, \aagnone\ and \aagntwo\ against their true values, for \textsc{IllustrisTNG}, \textsc{Simba}, \textsc{Swift-EAGLE} and \textsc{Astrid}.
As input data we provide all luminosity functions and colour distributions described in \sec{architecture} at a single redshift, $z = 0.1$.
We also calculate a number of statistics describing the goodness of fit, including the Root Mean Squared Error (RMSE),
\begin{align}
    \mathrm{RMSE} = \sqrt{ \frac{1}{N} \sum^{N}_{i=1} (\theta_i - \mu_i)^2} \;,
\end{align}
where $\theta_i$ are the true values for simulation $i$, $\mu_i$ are the mean predicted values, and $N$ is the number of samples. We also calculate the mean relative error,
\begin{align}
    \epsilon = \frac{1}{N} \sum^{N}_{i=1} \dfrac{| \theta_i - \mu_i |}{\mu_i} \;.
\end{align}
Both these statistics measure the \textit{accuracy} of the estimate (lower values being more accurate).
We also calculate the coefficient of determination,
\begin{align}
    R^2 = 1 - \dfrac{ \sum^{N}_{i=1} (\theta_i - \mu_i)^2 }{ \sum^{N}_{i=1} (\theta_i - \hat{\theta})^2 } \;,
\end{align}
where $\hat{\theta}_i$ are the mean values of the true parameters. 
This measures both the precision and the \textit{accuracy} of the estimate (values close to 1 being more precise \textit{and} accurate).
Finally, we measure the reduced chi-squared,
\begin{align}
    \chi^2 = \frac{1}{N} \sum^{N}_{i=1} \left( \frac{ (\theta_i - \mu_i) }{ \sigma_i} \right)^2 \;,
\end{align}
where $\sigma_i$ are the estimated 1$\sigma$uncertainties for simulation $i$.
This measures the \textit{accuracy of the estimated posterior uncertainties}; values close to 1 indicate that the magnitude of the errors (approximated by the marginal standard deviation $\sigma$) has been estimated correctly, whereas values above / below 1 indicate an under- / over-prediction of the error.
For all galaxy formation models and parameters (except for a couple of exceptions discussed below) the reduced chi-squared value lies between 0.76 and 1.55, which indicates that all errors are well estimated.

There are some clearly visible trends in Figures \ref{fig:single_predictions_A} \& \ref{fig:single_predictions_B}.
\omegam\ is predicted precisely and accurately across all simulations, with a RMSE of 0.03 and $R^2$ of 0.93 in Swift-EAGLE, though slightly less well in Astrid (RMSE=0.07, $R^2$=0.63).
This agrees well with the large variations in the LFs and colour distributions as a function of \omegam\ seen in \sec{1P_omegam}.
The value of \sigmae\ is also predicted, somewhat surprisingly; all we have provided as observed data are luminosity functions and colour distributions, which do not include any explicit spatial information from which to derive clustering constraints.
The precision and accuracy strongly depend on the galaxy formation model, with the best precision and accuracy found in IllustrisTNG (RMSE=0.05, $R^2$=0.81), and the worst of both in Astrid (RMSE=0.09, $R^2$=0.38).
This suggests that specifics of the galaxy formation model affect how \sigmae\ impacts the population galaxy emission, as seen in \sec{1P_sigmae}, highlighting the importance of taking into account such specifics when marginalising over astrophysics in a cosmological analyses using such observable features.

\begin{figure*}
    \includegraphics[width=\textwidth]{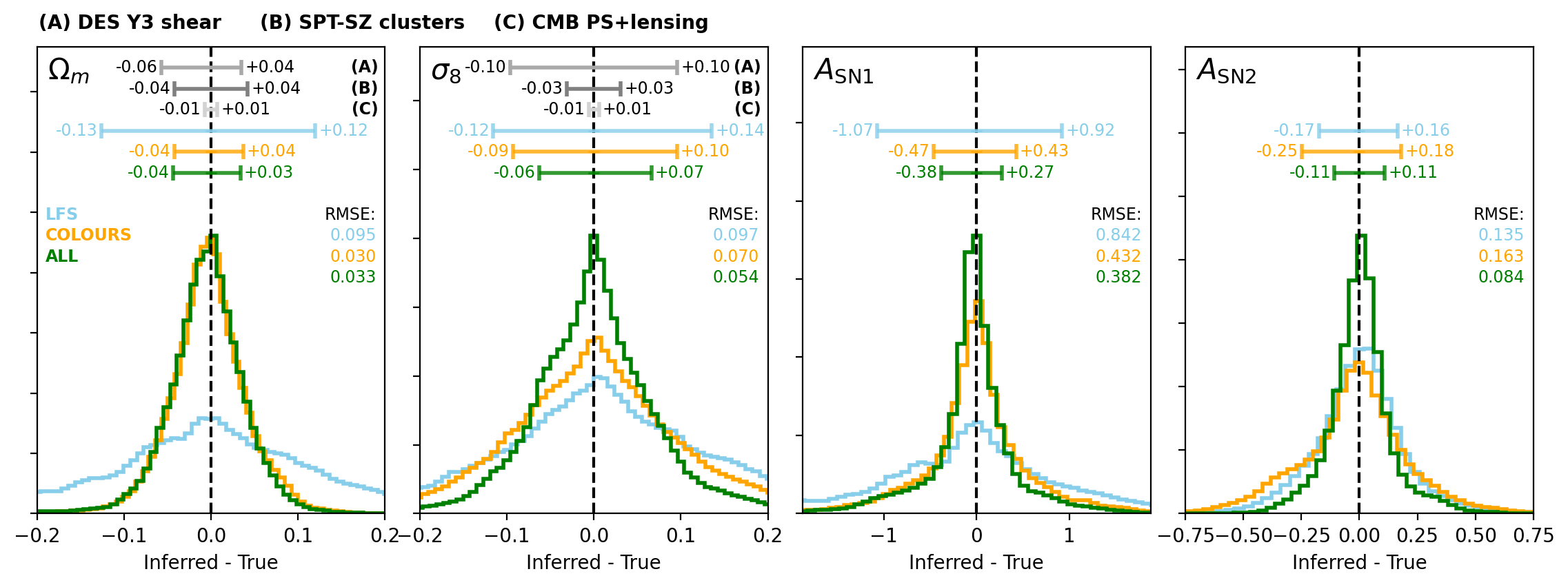}
    \includegraphics[width=\textwidth]{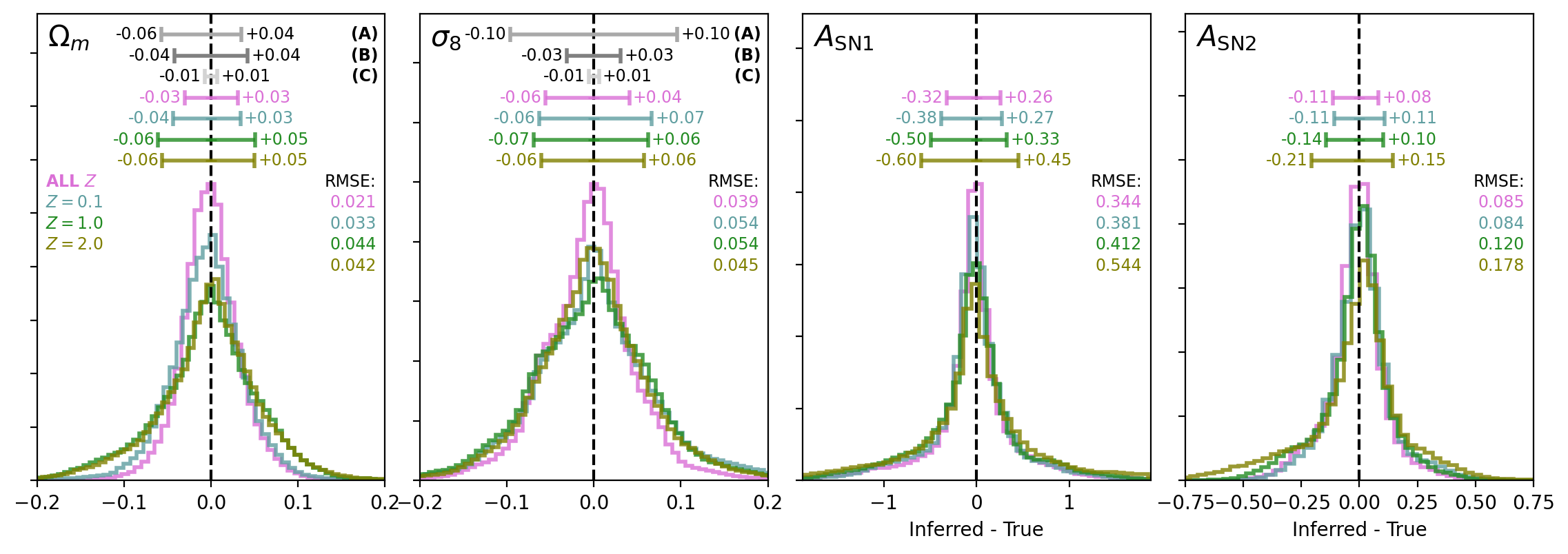}
    \caption{
    Combined marginal posterior distribution of \omegam, \sigmae, \asnone and \asntwo\ from all test set objects in the \textsc{IllustrisTNG} suite. \textit{Top row:} models trained using luminosity functions (blue), colours (orange), or both combined (green) at $z = 0.1$. \textit{Bottom row:} models trained using luminosity functions and colours combined at $z = 0.1$ (navy), $z = 1.0$ (green), $z = 2.0$ (olive), and all three redshifts combined (pink).
    The median (across the test set) 16$^{\rm th}$ -- 84$^{\rm th}$ percentile range of the marginal posteriors is indicated by the bars at the top of each subplot; the same marginal constraints from the Dark Energy Survey (DES) Year 3 angular power spectra cosmic shear analysis \citep[A;][]{doux_dark_2022}, the South Pole Telescope (SPT) Sunyaev-Zel'dovich selected cluster abundance analysis \citep[B;][]{de_haan_cosmological_2016}, and the Planck 2018 CMB analysis \citep[C;][including the lensing reconstruction]{planck_collaboration_planck_2020}, are also shown for comparison.
    }
    \label{fig:feature_importance}
\end{figure*}

In addition to these goodness of fit metrics, we also look at the median of the 16$^{\rm th}$ -- 84$^{\rm th}$ percentile range of the posterior distributions across the test set, a measure of the accuracy of the posterior predictions.
This summary statistic can then be directly compared to those obtained from other observational cosmological probes.
We compare to the constraints from three different experiments: the Planck 2018 CMB constraints including the lensing reconstruction \citep{planck_collaboration_planck_2020}, the Dark Energy Survey (DES) Year 3 angular power spectra cosmic shear analysis \citep{doux_dark_2022}, and the South Pole Telescope (SPT) Sunyaev-Zel'dovich selected cluster abundance constraints \citep{de_haan_cosmological_2016}.
The marginal constraints for \omegam\ and \sigmae\ from these experiments are shown in Figures \ref{fig:single_predictions_A} \& \ref{fig:single_predictions_B}.
We also present the median 16$^{\rm th}$ -- 84$^{\rm th}$ percentile range of the marginal posterior distributions across the test set for these parameters in \fig{feature_importance} for IllustrisTNG, alongside the observational constraints (other models are shown in \app{feature_importance_models}).
We emphasise that we are only comparing the \textit{precision} of our estimates, and not the accuracy, since we are not applying our pipeline to actual observational data.
Our constraints are clearly not competitive with those obtained from Planck CMB measurements.
However, we achieve comparable constraints on \omegam\ to those from DES and SPT, and for \sigmae\ our constraints are more precise than those from DES, but less precise than those from SPT.
We emphasise that even where these constraints are not competitive, they are still complementary to these standard probes.

As discussed in Sections \ref{sec:1P_sn} \& \ref{sec:1P_agn}, the feedback parameters are not directly comparable between galaxy formation models.
However, in general, the stellar feedback parameters are all constrained to some degree across all models.
Certain parameters, such as \asntwo\ for IllustrisTNG, are predicted remarkably well (RMSE=0.09, $R^2$=0.96), driven by the clear impact on the UV and optical LFs and colour distributions.
In contrast, the AGN parameters are, in general, not constrained at all, tending to just predict the median of each distribution.
There are some exceptions, such as \aagnone\ for Swift-EAGLE (RMSE=0.54, $R^2$=0.66) and \aagntwo\ for Astrid (RMSE=0.76, $R^2$=0.44).
The fact that these both control the thermal feedback mode suggests that this feedback mode may have more of an impact on the stellar contents of relatively lower mass haloes, present in these CAMELS simulation volumes.
However, the discussion in \sec{1P_agn} clearly shows how these parameters affect galaxy growth, and the resultant observables, in opposite directions; in Swift-EAGLE, increasing the efficacy of the feedback by dumping more energy inhibits galaxy growth more efficiently, whereas in Astrid the opposite is true, as increasing the thermal energy per unit accreted mass curtails SMBH growth, preventing SMBHs from growing massive enough to start the more effective jet feedback channel.
This may also explain why, despite the large differences seen in the LFs and colours for Astrid, there are significant degeneracies between parameters, which degrades the posterior prediction accuracy compared to e.g. \asntwo. 

\subsection{Feature Importance and Redshift Dependence}
\label{sec:feature_importance}

Given the promising constraints on astrophysical and cosmological parameters demonstrated in \sec{single_sim}, we now ask what particular information in our modelled data (or features) is providing constraints on these parameters.
We also explore how using information at different redshifts affects our inference, and whether combining distribution functions from multiple redshifts can improve parameter constraints.

\begin{figure*}
	\includegraphics[width=0.49\textwidth]{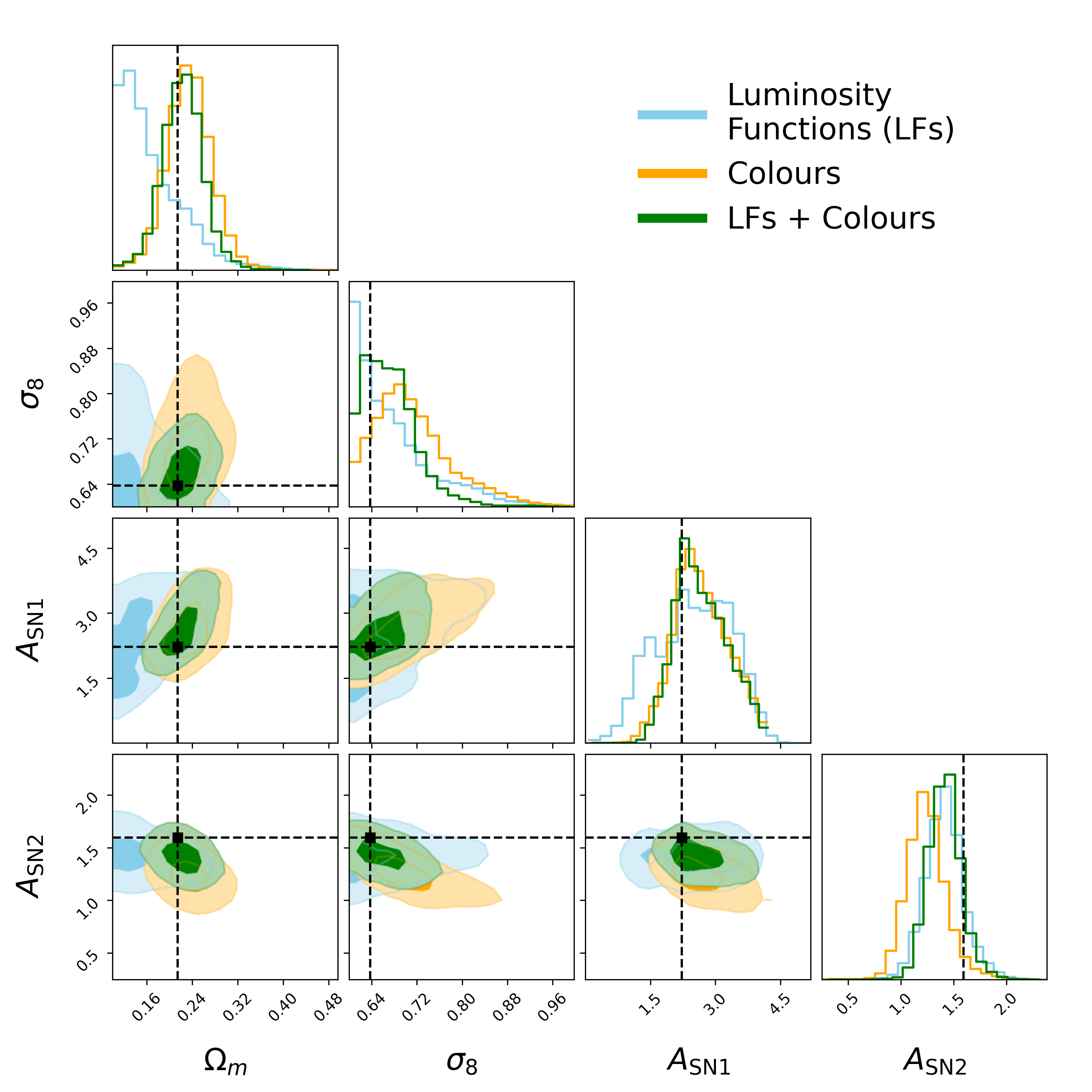}
    \includegraphics[width=0.49\textwidth]{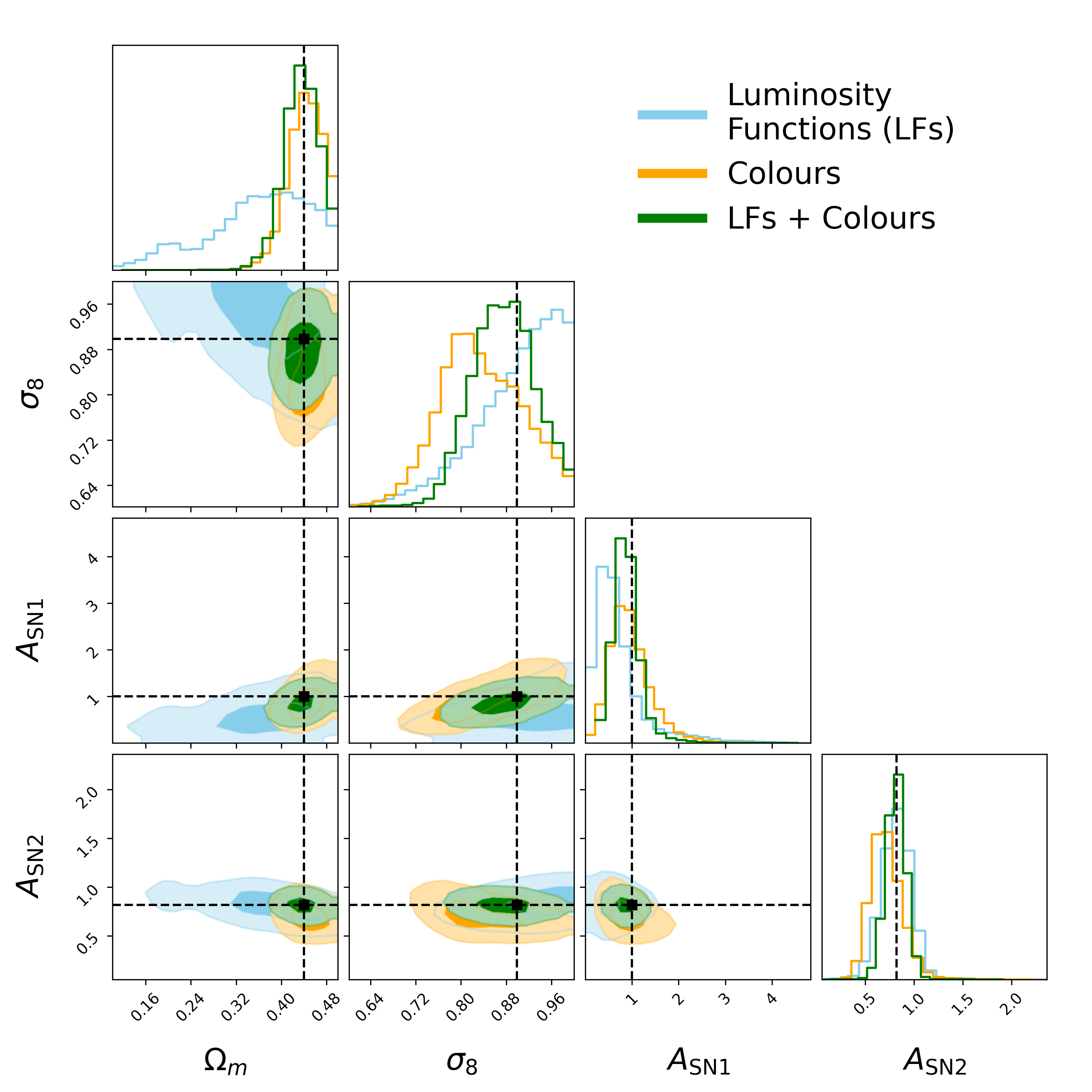}
    \includegraphics[width=0.49\textwidth]{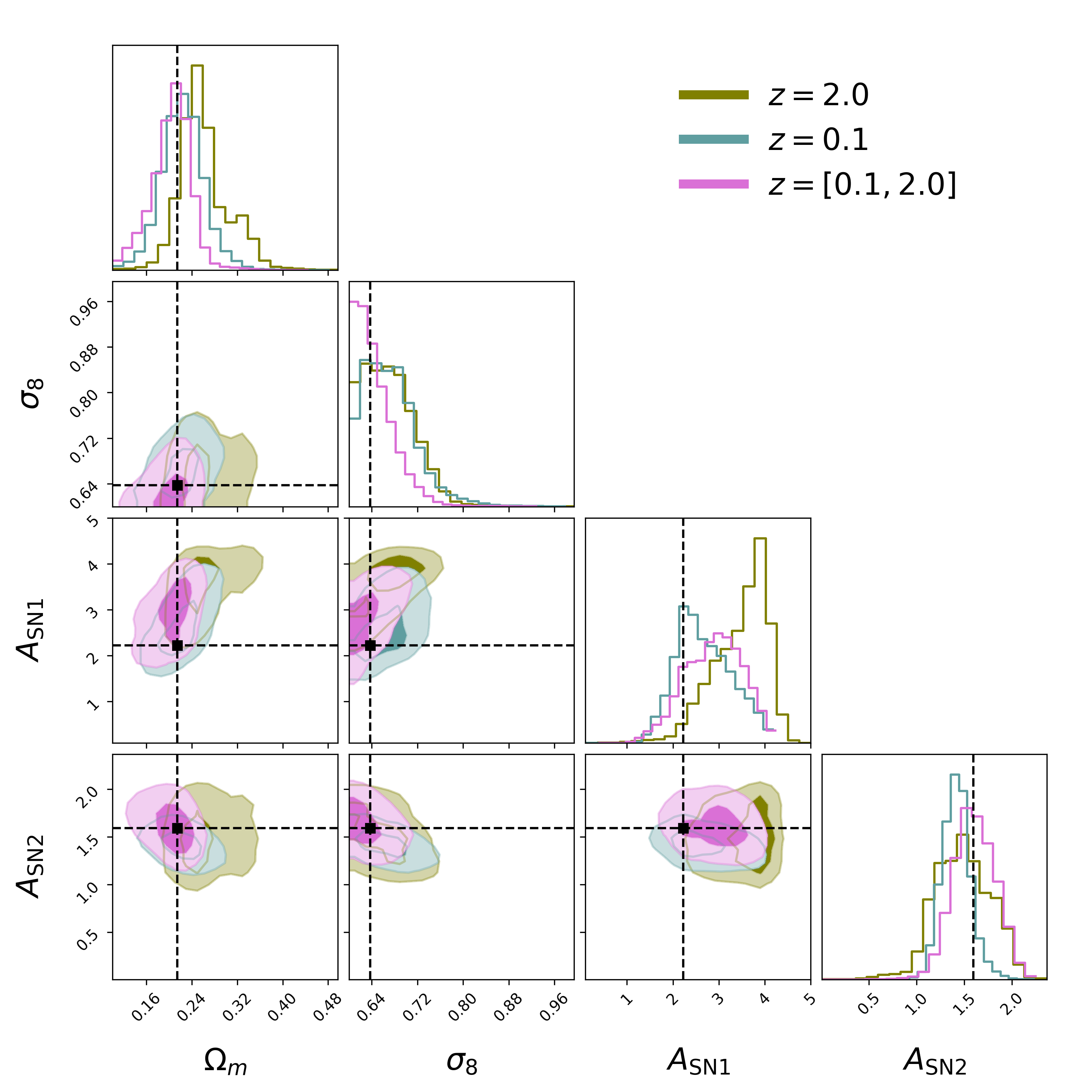}
    \includegraphics[width=0.49\textwidth]{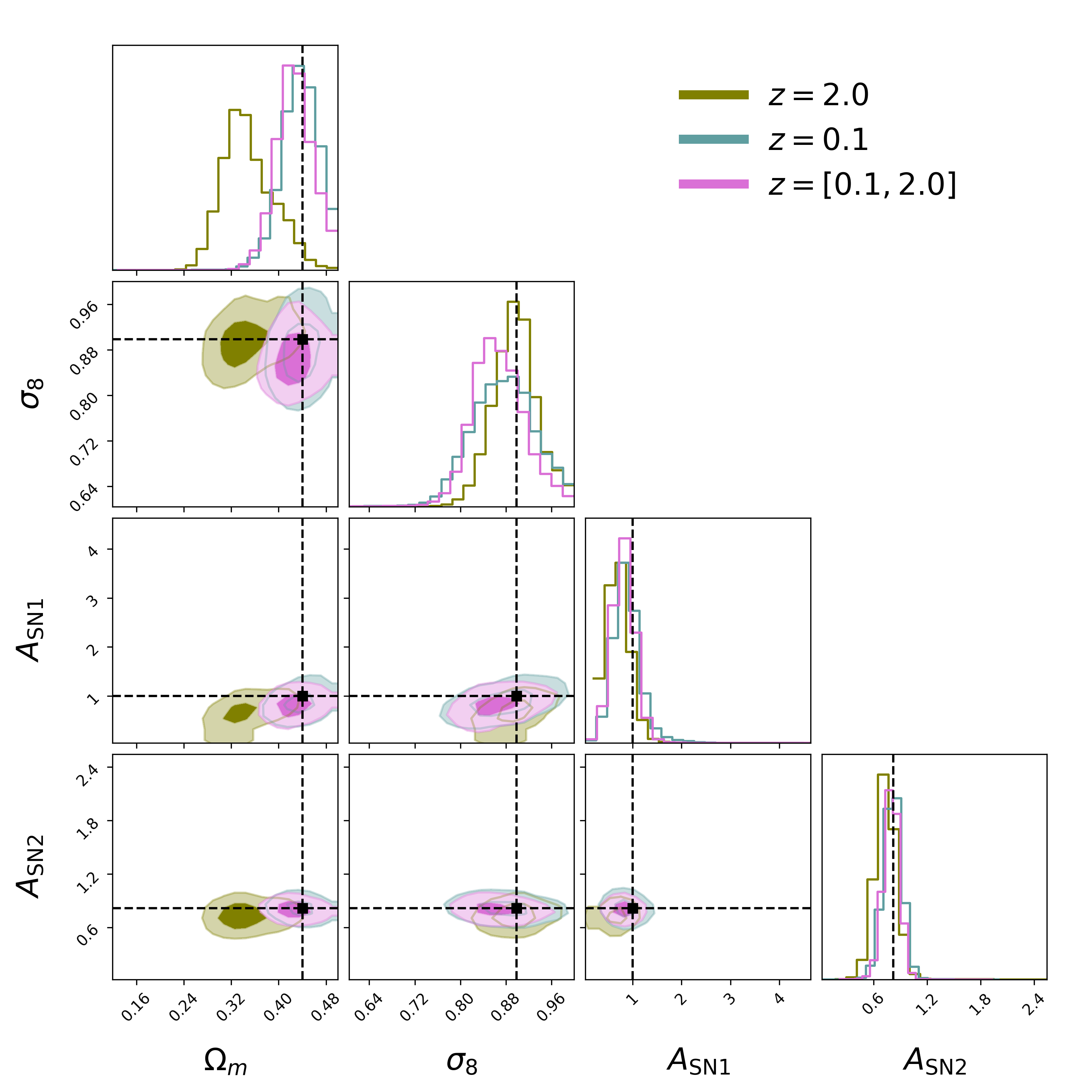}
    \caption{Posterior 2D and 1D marginal posteriors for two randomly selected test set simulations from the IllustrisTNG suite.
    Each column of corner plots shows the same test set simulation posteriors, assuming different input features.
    The true values are indicated by the black dashed lines.
    \textit{Top row:} posteriors given luminosity functions (blue), colours (orange), or both combined (green). \textit{Bottom row:} posteriors given luminosity functions and colours at $z = 0.1$ (navy), $z = 2.0$ (olive), and both redshifts combined (pink).}
    \label{fig:TNG_posterior_examples}
\end{figure*}

\subsubsection{Feature Importance}
To begin with we show four examples of the posterior constraints on simulations from the \textsc{IllustrisTNG} test set at $z = 0.1$ as 2D corner plots (\fig{TNG_posterior_examples}).
The left column shows the constraints achieved when providing luminosity functions (LFs) only, colour distributions only, or the combination of LFs and colours.
These clearly show the impact of the different features on the constraints for different parameters in these specific examples, and the correlations between the constraints on different parameters in some cases.

To better summarise the impact of the different features across the whole test set, we combine the posteriors in \fig{feature_importance} by binning the stacked residuals; where these are more peaked at zero, the tighter the constraints on this parameter.
We also show the median 16$^{\rm th}$ -- 84$^{\rm th}$ percentile range of the marginal posterior distributions across the test set, to summarise the precision of our estimates.
Here we can clearly see the importance of the various features on each parameter.
Interestingly, it appears that LFs do not place tight constraints on \omegam.
The variation with \omegam\ seen in \fig{1P_omegam} in the FUV and $r$-band is mostly for low values of \omegam\ (< 0.3); at higher values the differences in the LFs are much smaller, which introduces large degeneracies, and leads to inflated error estimates across the test set.
Colours, on the other hand, do lead to very tight constraints on \omegam; the variation can be seen across the range of \omegam\ in \fig{1P_omegam}, which explains the low errors across the whole test set.
For \sigmae\ it appears that the use of LFs and colours individually give relatively similar constraints, but combined they provide much tighter constraints.
For \asnone\ and \asntwo\ it is a similar story; comparable constraints for LFs and colours individually, but tighter constraints combined.

The fact that colours drive the tight constraints on \omegam\ is a somewhat surprising result.
\cite{jo_calibrating_2023} show how using just the binned galaxy stellar mass function can achieve tight constraints on cosmological and astrophysical parameters, even in the presence of simulation uncertainty.
We do not include the $K$-band in this analysis, which is known to closely trace the underlying stellar mass, however one might naively expect the optical luminosity functions to still provide sufficient information on the total stellar mass content.
This suggests that the impact of age and metallicity degeneracies may play an important role.
It also highlights how the colours trace the stellar age of the global galaxy population, which can be significantly higher in a universe with a lower \omegam\ (up to 18 Gyr for \omegam=0.1), since low mass stars have exceedingly long lifetimes \citep[exceeding 100 Gyrs;][]{laughlin_end_1997}.

\subsubsection{Redshift Dependence}

\fig{TNG_posterior_examples} also shows two example corner plots (bottom row) when using the combination of the LFs and colours obtained at different redshifts.
In all these examples the marginal posteriors are tighter when combining information from different redshifts, as expected.

We summarise results from the whole test set in \fig{feature_importance}, as we did for the LFs and colours.
For \omegam, using data at $z = 0.1$ has the most constraining power, and this degrades at higher redshift.
However, the improvement from combining information from the three redshifts is significant, decreasing the RMSE by >0.01.
Interestingly, for \sigmae\ the most constraining power is provided by the distribution functions at $z = 2$; this suggests that the impact of matter clustering on the luminosities and colours of galaxy populations may be more evident at cosmic noon.
This could be explained by the larger impact of \sigmae\ on the halo mass function at higher redshift \citep{tinker_toward_2008}, so the information content from the LFs may be higher than at $z = 0.1$, where it is dominated by the colours.

We did not combine higher redshift snapshots due to the low number of galaxies in the CAMELS volumes at $z > 2$. This led to our distributions becoming increasingly noisy, and led to particularly poor sampling speeds from the NPE where most bins were empty.
This highlights a known issue with SBI approaches when handling missing data \citep[e.g.][]{wang_sbi_2023}; a number of modern methods are being developed to address this \citep[e.g. \textsc{Simformer};][]{gloeckler_all--one_2024}.

It is worth noting that, in these examples, by combining snapshots at different redshifts we are at risk of using the same galaxies at different points in their evolution.
This may bias our constraints compared to a real survey, since galaxies at different redshifts along the lightcone will not be associated in the real Universe, and this evolution encoded at different times may be more constraining than independent measurements.
In order to overcome this issue we could subdivide each simulation volume and take galaxies from subboxes at different redshifts, however for the CAMELS boxes considered here the volumes are too small to get sufficient statistics.
We will more robustly assess the impact of these correlations in future work with the larger volume CAMELS-SAM simulations \citep{perez_constraining_2023}.

\subsection{Generalization Across Different Models}
\label{sec:robustness}

\begin{figure*}
	\includegraphics[width=\textwidth]{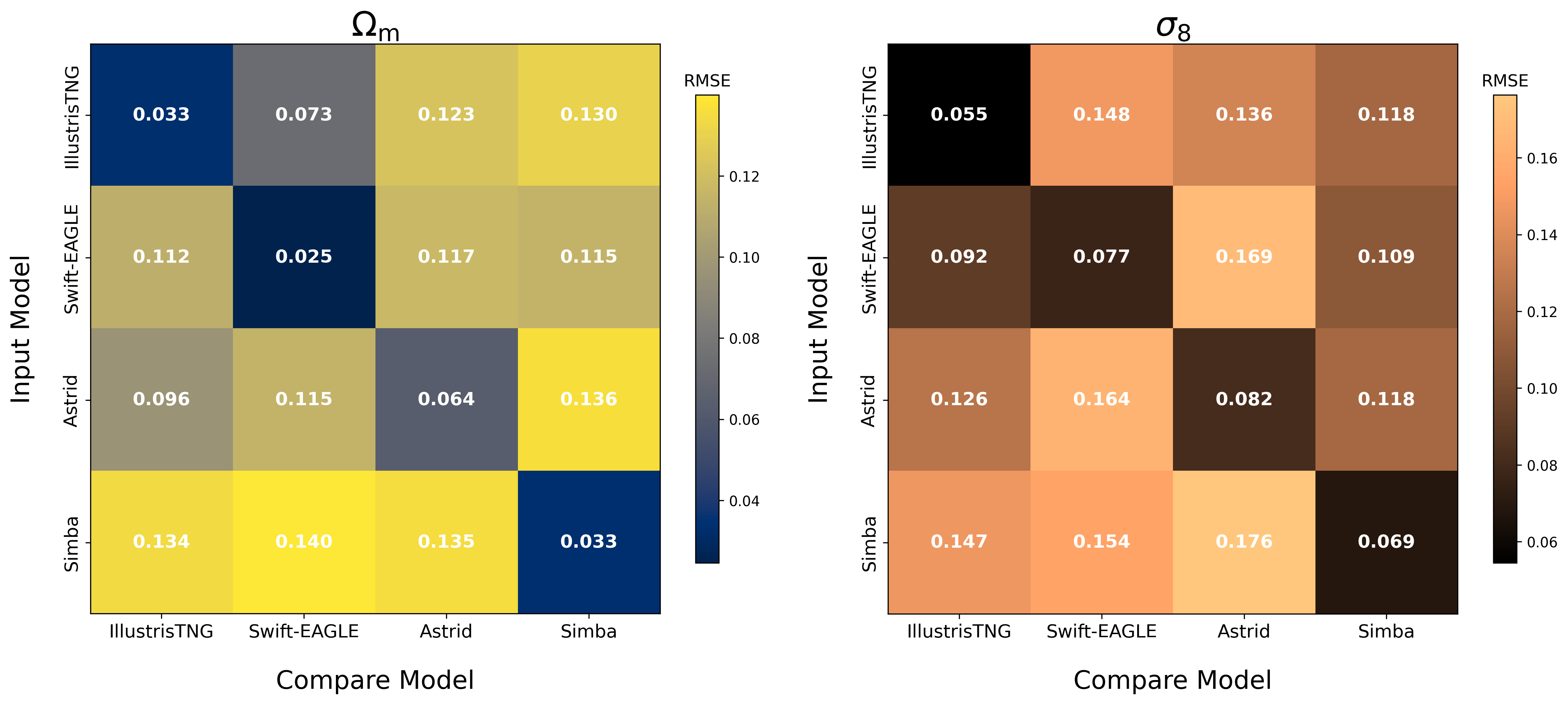}
    \caption{
        RMSE for the posterior predictions from a given input model when applied to feature vectors from another model, for the cosmological parameters \omegam\ and \sigmae.
        Where the input model and comparison model are the same, we limit the comparison to the test set simulations.
    }
    \label{fig:intersim_predictions}
\end{figure*}

One of the motivations for including multiple galaxy formation simulations in the CAMELS suite was to enable tests of the robustness of trained machine learning algorithms to different training simulations \citep{de_santi_robust_2023,ni_camels_2023}.
In this case, we can use the forward modelled photometry from one simulation as input to our SBI pipeline trained on another, and test the recovery of the underlying parameters.
Since the astrophysical parameters represent different aspects of each model we cannot perform inference on these between models, so instead we focus on the cosmological parameters \omegam\ and \sigmae.

\fig{intersim_predictions} shows the RMSE of the posterior predictions for these parameters when using an input model applied to data from a comparison model.
We find relatively poor recovery of parameters when testing between simulations, and the precision is dependent on the model considered and the data it is applied to.
For example, for \omegam\ the IllustrisTNG model applied to Swift-EAGLE data performs relatively well, with an RMSE of 0.7, comparable to the within-simulation Astrid model results.
However, the Simba model performs poorly when applied to all other simulation suites for \omegam.
However, the trends seen for \omegam\ do not necessarily translate into precise estimates for \sigmae; the IllustrisTNG model performs most poorly when applied to Swift-EAGLE for this parameter compared to the other suites.
The accuracy of the inter-simulation predictions is also generally poor, with $R^2$ values below 0.5 for all models and parameters.

These results are similar to those found in other CAMELS studies \citep[e.g.][]{de_santi_robust_2023,de_santi_field-level_2025,ono_debiasing_2024}.
The reason for this relatively poor recovery of the parameters for the intersim tests in this study can be clearly seen in Figures \ref{fig:1P_omegam}, \ref{fig:1P_sigmae}, \ref{fig:1P_asnone}, \ref{fig:1P_asntwo}, \ref{fig:1P_aagnone} and \ref{fig:1P_aagntwo}, which show the range of luminosity functions and colour distributions across the 1P sets of each subgrid model.
There is very little agreement or overlap in the detailed distributions, which leads to out-of-distribution errors when performing inference.
We have tested using just luminosity functions or just colour distributions, as well as reducing the fidelity of the distribution functions by reducing the number of bins, and achieve similar results.

The main source of this lack of robustness is the very different subgrid prescriptions in each model, which lead to different distributions of point-in-time properties, such as stellar mass and star formation rate, as well as different overall star formation histories (Iyer et al. \textit{in prep.}).
Another source of inflexibility is in our forward model for galaxy emission, which assumes a simple dust prescription, and fixes many key parameters, such as the nebular cloud dispersion time.
In future work we will self-consistently modify elements and parameters of the forward model, such as the dust attenuation model, which should lead to increased overlap in colour and luminosity space between models.
We will also explore contrastive learning \citep{le-khac_contrastive_2020} and domain adaptation approaches \citep{roncoli_domain_2023,ciprijanovic_domain_2020,ciprijanovic_semi-supervised_2022}, which have shown promise in overcoming these issues when combining models, and for extracting domain-invariant features, as well as approaches for excluding significant outliers \citep{echeverri-rojas_cosmology_2023,de_santi_field-level_2025}.

\section{Discussion}
\label{sec:discussion}

\subsection{Cosmological Constraints}


We have found that, by just considering UV-optical luminosity functions and colours, we can obtain precise and accurate estimates of \omegam\ across different subgrid galaxy evolution models.
In terms of accuracy, for IllustrisTNG, Swift-EAGLE and Simba we obtain a RMSE across the test set of 0.03, and for Astrid slightly higher at 0.07.
With regards to precision, we obtain a 16$^{\rm th}$ -- 84$^{\rm th}$ percentile range of the marginal posterior distribution of \omegam\ of 0.07 for IllustrisTNG \& Swift-EAGLE, 0.09 for Simba, and 0.16 for Astrid.
These are more accurate than those obtained by \cite{hahn_cosmology_2024} on galaxies from the NASA-Sloan Atlas, despite only using summary statistics measured on a fraction of the number of galaxies.
This compares favourably with constraints from observational experiments, such as the Dark Energy Survey (DES) Year 3 angular power spectra cosmic shear analysis \citep[0.10;][]{doux_dark_2022}, and the South Pole Telescope (SPT) Sunyaev-Zel'dovich selected cluster abundance analysis \citep[0.08][]{de_haan_cosmological_2016}, though it is not competitive with the constraints from the Planck 2018 CMB analysis where the lensing reconstruction is considered \citep[0.02;][]{planck_collaboration_planck_2020}.
In \sec{feature_importance} we show that, in IllustrisTNG, it is predominantly galaxy colours that drive these constraints on \omegam.

This suggests that it is \textit{internal} galaxy properties that are affected by changes in the matter density, not just the overall abundance of haloes and therefore galaxies.
This is supported by the dependence of the ML ratios on \omegam, which implicitly factors out the change in halo abundance.
These findings somewhat support what was found in \cite{villaescusa-navarro_cosmology_2022} and \cite{echeverri-rojas_cosmology_2023}, where the properties of a single galaxy were used to obtain significant constraints on \omegam.

We also obtain constraints on \sigmae, describing the level of matter clustering, presented in \sec{single_sim}, despite not including any spatial information in our feature vector.
For all four simulations we achieve a RMSE $\leqslant 0.09$ on \sigmae, and a 16$^{\rm th}$ -- 84$^{\rm th}$ percentile range of the marginal posterior distribution of 0.13 for IllustrisTNG, and $\sim$0.18 for Swift-EAGLE, Simba \& Astrid.
In \sec{feature_importance} we show that it is a combination of galaxy colours and luminosity functions that drives this good predictive accuracy in IllustrisTNG, which is in turn driven by the older ages of galaxy stellar populations in simulations with a higher \sigmae.
\sigmae\ also affects the shape and normalisation of the halo mass function, particularly at the bright end, which would also impact the optical-UV luminosity functions.

We stress that when interpreting these constraints quantitatively we must keep in mind that this analysis does not include any observational uncertainties, which will degrade our constraints \citep[see e.g.][]{hernandez-martinez_cosmological_2025}.
We discuss other observational considerations below.
Additionally, we do not include any estimate of the simulation uncertainty; \cite{jo_calibrating_2023} show that this can be derived from the CV set and used within an emulator to assess its impact, and find that it can lead to wider estimated posteriors.
Finally, our dust model does not depend in any way on galaxy properties.
Including the correlations and covariances between, for example, the mass of dust and the optical depth, or the star-dust geometry and the form of the attenuation curve, will lead to degeneracies with some of the effects on the LFs and colours seen in Figures \ref{fig:1P_omegam} and \ref{fig:1P_sigmae}, which will degrade the constraints on cosmological parameters.
However, conversely, the dust model itself may leak cosmological information, due to dependencies on the SFZH from each galaxy; we explore this effect with our current galaxy-independent dust model briefly in \sec{intrinsic}.
We will explore a more sophisticated model that takes these correlations into account in future work.

\subsection{The Impact of Mass and Local Environment, and the Limitations of Volume and Resolution}
\label{sec:mass_env}

The mass and local environment of a galaxy can have significant effects on its observable properties, particularly its colour \citep{peng_mass_2012}.
More massive galaxies tend to be older and redder, due to their extended star formation histories, and feedback from central black holes \citep{di_matteo_energy_2005}.
Lower mass galaxies tend to be bluer and more highly star forming, unless they are satellites, in which case they experience additional quenching mechanisms that lead to redder colours \citep{hogg_dependence_2004,van_den_bosch_importance_2008,weinmann_environmental_2009}.
The trends have been well studied in the literature, and are often referred to as the \textit{colour-magnitude} (where magnitude is assumed to correlate with mass to first order) and \textit{colour-environment} relations.
It is therefore of interest to understand how these subsets of the galaxy population impact our luminosity functions and colours, and propogate into our constraints on cosmological and astrophysical parameters.

However, we must consider the limitations introduced by the volume and resolution of our simulations on our ability to simulate representative populations of these galaxies.
The CAMELS simulations are relatively small in volume, and have relatively low mass resolution.
This means that the number of high mass haloes is limited, due to the statistically small volume, but also the lack of large scale modes of the power spectrum that can be captured in small periodic volumes.
This also reduces the number of high mass host haloes for lower mass satellite galaxies.
In a similar vein, the resolution of the simulations limits the ability to resolve smaller scale structures, which further restricts the satellite populations.
Despite this, we see that, for all galaxies with stellar mass $>10^{8} \, \text{M}_\odot$, less than 50\% are satellite galaxies across all models considered, which is approximately similar to the fractions found in larger volume simulations.
However, we find that < 10\% of all galaxies have stellar masses $>10^{10} \, \text{M}_\odot$ across all models, and the fraction of satellites for these higher mass galaxies are lower, $< 40\%$.
These effects may explain the poor numerical convergence of colour distributions with resolution seen in other studies \citep[e.g.][]{trayford_colours_2015,nelson_first_2018}.

We have checked our colour distributions and LFs for satellite galaxies and centrals, and found no appreciable difference in the distributions if central galaxies are excluded.
Centrals tend to contribute a greater fraction of the brightest galaxies across all models and bands, and satellites to the fainter number densities.
Satellites tend to have marginally redder colour distributions across all suites, but this effect is not large; the red peak in Swift-EAGLE for $r-i$, for example, is shifted by $\sim +0.01$ in this colour space.
This may be due to the limited resolution in CAMELS discussed above; many environmental effects, such as tidal and ram-pressure stripping, may not be properly captured.
We also checked the impact of high mass galaxies, and found more interesting trends.
High mass galaxies in all models dominate the number densities of the brightest galaxies, and therefore contribute significantly to these bins in our LFs.
In IllustrisTNG and Astrid massive galaxies also contribute significantly to the red galaxy populations.
In Swift-EAGLE and Simba the colour distributions of high and low mass galaxies are more similar.
This is particularly interesting given the bimodal distribution seen in IllustrisTNG and Swift-EAGLE, compared to the unimodal distribution in Simba and Astrid.

\subsection{Performing SBI directly on observations}
\label{sec:difficulties}

When performing inference in a simulation based inference framework on \textit{cosmological} parameters in particular, one must take into account the impact of changing the cosmological parameter on any forward modelled properties or distribution functions.
In this study, where we focus on volume normalised rest-frame distribution functions, this manifests in two main ways; $K$-corrections for rest-frame photometry, and cosmological volume normalisation.

For rest-frame photometry, the change in luminosity distance implied by a change in \omegam\ leads to differences in the redshift, and therefore a change in the rest-frame part of the spectrum probed by the photometric filter (at a given redshift).
In observed catalogues, the emission within a filter bandpass in the rest-frame is estimated by assuming a $K$-correction \citep{hogg_k_2002}.
This correction assumes a shape for the spectrum, which depends on the assumed galaxy population, and its evolutionary history.
These properties are explicitly cosmology dependent; if the cosmology is updated, the $K$-correction (or, more specifically, the $E$-correction) must be updated in line with this cosmology.
In an SBI framework, updating the target observable on each parameter iteration is not feasible.
Instead, one could forward model into \textit{observer-frame} fluxes, to avoid the application of a $K$-correction altogether.

Similarly, the change in apparent distance implied by a change in \omegam\ also leads to changes in the differential comoving volume at a given redshift. 
This impacts the volume normalisation of any derived number count distributions, such as luminosity functions.
In order to perform a like-for-like comparison, it is necessary to correct for this volume normalisation offset as a function of cosmology.
Alternatively, one can forward model into the observer-frame, and normalise by the sky area, to avoid any dependence on cosmological parameters.

In this study we focus on \textit{rest-frame}, \textit{volume normalised} distribution functions, rather than observer frame fluxes.
This is to better understand the impact of various modelling assumptions on the rest-frame properties, and better link to the underlying physics of each model.
We are also limited by the relatively small volume of the CAMELS simulations, which makes it more difficult to project into observer-space through the construction of a lightcone.

Finally, whilst observed luminosity functions from SDSS and GAMA are reasonably well covered over our 1P parameters in the intermediate magnitude range (see Figures \ref{fig:1P_omegam} -- \ref{fig:1P_aagntwo}), it is clear that we struggle to reproduce the detailed, complex colour distributions.
Whilst the broad trends are captured, some bins in colour are out of distribution at this resolution.
This is due to the known difficulty of replicating the detailed abundance and colour distributions in observational space, even over this extended parameter space \citep[see \textit{e.g.}][]{trayford_colours_2015}.
In future work we will investigate the impact of varying more parameters in our forward model, such as the dust optical depth and attenuation curve, to investigate whether this increases the compatibility of observed colours with the simulations.

For these reasons, we avoid performing inference directly on the observed distribution functions in the rest-frame, and leave a detailed comparison with observations in the observer-frame to future work.

\subsection{Forward Modelling Limitations}
\label{sec:limitations}

As touched on briefly in Sections \ref{sec:forward_model} and \ref{sec:camels}, there are a number of limitations in our modelling of the emission from galaxies, many of which are set by the fidelity of the fiducial CAMELS suites (see also \sec{mass_env}).
Due to the low spatial and mass resolution of the hydro simulations we do not resolve galaxy sizes, which precludes their inclusion as features.
We also therefore cannot include features such as emission profiles, or resolved maps of the emission in different bands.
These features may contain crucial cosmological and astrophysical information, and will be regularly estimated in upcoming wide field surveys, such as those from Euclid and Rubin.

The mass resolution in particular also places constraints on resolved short timescale variations in the star formation history.
These are important for predicting the recent star formation. which is key for self-consistently producing the contribution from nebular line and continuum emission.
This recent star formation can be estimated by smoothing the star formation history, or even by adding small scale variation through power spectral density methods \citep{iyer_diversity_2020}, however these typically require training from higher resolution input simulations.
Finally, the application of detailed dust radiative transfer approaches, whilst also limited by their computational complexity, are also inappropriate for such low-resolution simulations, since the star-dust distribution will be poorly estimated.
In summary, whilst the fidelity of the CAMELS simulations is sufficient for an an analysis of broad band photometric emission in the UV-optical, higher resolution simulations will be required to explore these more detailed spatial and time-sensitive features, and for enabling more sophisticated forward modelling approaches.

\section{Conclusions}
\label{sec:conclusions}

We have forward modelled the UV-NIR emission from galaxies in the CAMELS simulation suites, and used these to perform Simulation Based Inference (SBI) on galaxy luminosity functions and colours to predict the values of the cosmological parameters \omegam\ and \sigmae, as well as two parameters controlling the strength of stellar feedback, and two controlling the strength of feedback from active galactic nuclei (AGN).
All photometric catalogues are made publicly available at \url{https://camels.readthedocs.io/en/latest/data_access.html} for the community to use.

Our findings are as follows:
\begin{enumerate}
    \item We model luminosity functions and colour distributions in the \textit{GALEX} FUV-NUV and SDSS \textit{ugriz} photometric bands for the \textsc{Swift-EAGLE}, \textsc{IllustrisTNG}, \textsc{Simba} and \textsc{Astrid} galaxy formation models from $z = 0.1$ to $6$, and find significant differences between their fiducial predictions, particularly in their optical and UV colours.
    
    \item Cosmological and astrophysical parameters significantly impact the UV and optical luminosity functions and colours, with the relative impact depending on the subgrid galaxy formation model. In particular, we find that \omegam\ and \sigmae\ are both correlated with UV-optical colours, leading to redder galaxy distributions.

    \item We build an SBI model using the LtU-ILI framework to predict the values of the cosmological and astrophysical parameters from our combined luminosity functions and colours, and achieve relatively accurate and precise constraints on \omegam, \sigmae, \asnone\ \& \asntwo\ across all galaxy formation models. The AGN parameters are poorly constrained due to limitations in the CAMELS simulation volumes restricting the abundance of massive haloes.

    \item We perform a feature importance analysis using the IllustrisTNG model, and find that the tight constraints on \omegam\ are surprisingly driven by the UV-optical colour information. Colour information is similarly important for \sigmae\ constraints.

    \item We investigate constraints from distribution functions at higher redshifts, and find that \sigmae\ is best constrained at cosmic noon ($z = 2$), rather than $z = 0.1$. The tightest constraints for all parameters are obtained by combining information from multiple redshifts.

    \item The constraints on \sigmae\ are surprising given the lack of any spatial or clustering information in our chosen features.
    We find that enhanced clustering, parametrised by higher \sigmae, leads to earlier forming and more metal enriched galaxies, which manifests through their star formation and metal enrichment histories, and subsequently their luminosities and colours.

    \item As a test of robustness we apply a model trained on one simulation to another, and find poor generalisability, as seen in many other CAMELS analyses. We attribute this to differences in the underlying subgrid prescriptions in the galaxy formation models, but also identify inflexibility in our forward model as a potential culprit.
\end{enumerate}

These results suggest that distributions of galaxy luminosities and colours across the UV-optical wavelength range contain not only a wealth of information on astrophysical processes, but also significant cosmological information content.
Combined with more traditional sources of cosmological information, such as galaxy clustering statistics, there is potential to use these properties to provide more stringent constraints, and potentially break parameter degeneracies between probes.
Calibration of subgrid models in hydrodynamic simulations of galaxy evolution is also an example where linking parameters to data in a simulation-driven Bayesian approach has advantages \citep{elliott_efficient_2021,jo_calibrating_2023,kugel_flamingo_2023}; in this study we show how this can be done directly in observer space, rather than relying on inferred physical parameters obtained through SED fitting.

However, we have also shown the sensitivity of our inference to the underlying galaxy formation model.
All of the models presented here have been calibrated to, and show very similar results for, key distribution functions such as the galaxy stellar mass function.
However, there are clear differences in other property distributions, including, as shown here, their luminosity functions and colour distributions.
This reflects the highly degenerate nature of galaxy formation and evolution.
Whilst we have discussed a number of approaches for overcoming this model misspecification in \sec{robustness}, in general improved and more flexible models will help to reduce the domain shift when applying to real data.
In addition, exploring larger parameter spaces with more physically motivated parameterisations of the astrophysics \citep[e.g. the 28 parameter CAMELS SOBOL set;][]{ni_camels_2023} will help to cover a larger volume of the astrophysical and cosmological parameter hypervolume.

In addition, the simple forward model for galaxy emission employed here can also be modified.
The fidelity could be increased, for example by using radiative transfer approaches \citep{camps_skirt_2015,narayanan_powderday_2021} to self-consistently produce a wider range of attenuation curves seen in the real Universe \citep{narayanan_theory_2018}, or alternatively a larger range of parameters in the simple models employed here could be explored.
This is one of the key development goals of \textsc{synthesizer} (Lovell et al. \textit{in prep.}, Roper et al. \textit{in prep.}): to allow a rapid exploration of uncertain parameters in the forward model.
This is still a huge data challenge, given the size of a simulation suite such as CAMELS ($>$200 million galaxies across all suites and snapshots).

As described in \sec{difficulties}, there are a number of difficulties to comparing directly with observed data in an SBI framework.
In future work we will overcome a number of these limitations, at which point the method will be directly applicable to a number of future surveys, such as Euclid \citep{euclid_collaboration_euclid_2025} and LSST \citep{ivezic_lsst_2019}, containing billions of sources.

\section*{Acknowledgements}
We wish to sincerely thank the reviewer, Annalisa Pillepich, for their detailed and comprehensive review which significantly improved the clarity and rigour of the manuscript.
We thank Viraj Pandya for providing the GAMA catalogues used to construct the observed rest-frame colour distributions.
We also thank Mat Page, Sotiria Fotopoulou, Shy Genel, Yongseok Jo, ChangHoon Hahn, Greg Bryan, Rachel Cochrane, Will Handley and Stephen Wilkins for useful discussions.
This work was supported by the Simons Collaboration on “Learning the Universe”.
This work used the DiRAC Memory Intensive service (Cosma7) at Durham University, managed by the Institute for Computational Cosmology on behalf of the STFC DiRAC HPC Facility (www.dirac.ac.uk). The DiRAC service at Durham was funded by BEIS, UKRI and STFC capital funding, Durham University and STFC operations grants. DiRAC is part of the UKRI Digital Research Infrastructure.
The data used in this work were, in part, hosted on equipment supported by the Scientific Computing Core at the Flatiron Institute, a division of the Simons Foundation.
This research made use of the Spanish Virtual Observatory (https://svo.cab.inta-csic.es) project funded by MCIN/AEI/10.13039/501100011033/ through grant PID2020-112949GB-I00.
TS acknowledges support by NSF grant AST-2421845 and NASA grants 22-ROMAN22-0055 and 22-ROMAN22-0013.
DAA acknowledges support from NSF grant AST-2108944 and CAREER award AST-2442788, NASA grant ATP23-0156, STScI JWST grants GO-01712.009-A, AR-04357.001-A, and AR-05366.005-A, an Alfred P. Sloan Research Fellowship, and Cottrell Scholar Award CS-CSA-2023-028 by the Research Corporation for Science Advancement.
AG is grateful for support from the Simons Foundation.

We list here the roles and contributions of the authors according to the Contributor Roles Taxonomy (CRediT)\footnote{\url{https://credit.niso.org/}}.
\textbf{Christopher C. Lovell}: Conceptualization, Data curation, Formal analysis, Investigation, Methodology, Software, Visualization, Writing - original draft, Writing - review \& editing.
\textbf{Tjitske Starkenburg}: Conceptualization, Investigation, Validation, Writing - Review \& Editing.
\textbf{Matthew Ho}: Methodology, Software, Investigation, Verification. 
\textbf{Daniel Angl\'{e}s-Alc\'{a}zar, Francisco Villaescusa-Navarro}: Resources, Data Curation, Writing - Review \& Editing.
\textbf{Kartheik Iyer, Rachel Somerville, Laura Sommovigo}: Conceptualization, Investigation, Writing - Review \& Editing.
\textbf{William J. Roper}: Methodology, Software, Writing - Review \& Editing.
\textbf{Alice E. Matthews}: Validation, Writing - Review \& Editing.
\textbf{Romeel Dav\'{e}, Austen Gabrielpillai}: Writing - review \& editing.

\section*{Data Availability}

The full CAMELS public data repository is available at \url{https://camels.readthedocs.io}.
This includes the photometric catalogues.
Scripts for training and running the inference pipeline are made available at \url{https://github.com/christopherlovell/camels_observational_catalogues}.
Download and installation instructions for \textsc{Synthesizer} are available at \url{https://synthesizer-project.github.io/}.
Download and installation instruction for the LtU-ILI package are provided at \url{https://github.com/maho3/ltu-ili/tree/main}.



\bibliographystyle{mnras}
\bibliography{camels_obs}



\appendix

\section{Photometry database}
\label{sec:photo_database}

The photometric catalogues produced in this work are available at \url{https://camels.readthedocs.io/en/latest/data_access.html}.
We provide AB rest- and observer-frame magnitudes in filters from the GALEX ($FUV$ and $NUV$), GAMA ($ugrizYJHK$), Johnson ($UBVJ$), HST ACS (F435W, F606W, F775W, F814W, and F850LP), HST WFC3 (F098M, F105W, F110W, F125W, F140W and F160W), and JWST NIRCam (F070W, F090W, F115W, F150W, F200W, F277W, F356W and F444W) instruments.
These are provided for both the BPASS and BC03 SPS models, and for intrinsic and dust attenuated emission.
Full details on the forward modelling pipeline are provided in \sec{forward_model}.

\section{Coverage tests and Calibration}
\label{sec:coverage}

\begin{figure}
    \includegraphics[width=\columnwidth]{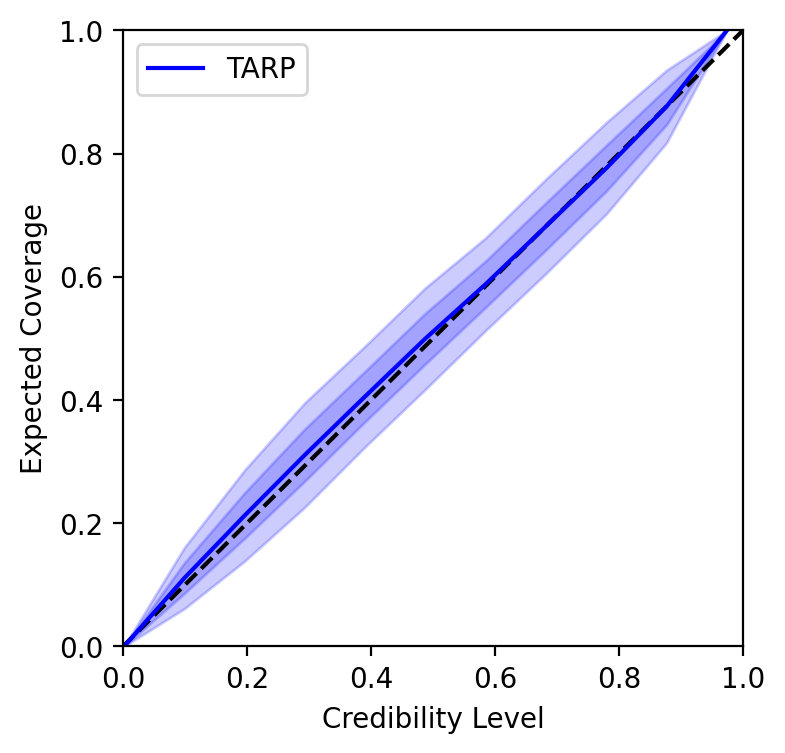}
    \caption{TARP coverage for the fiducial IllustrisTNG model, including dust attenuated luminosity functions and colours, at $z = 0.1$.}
    \label{fig:TNG_TARP}
\end{figure}

We carried out a number of tests of the model coverage for each of our trained SBI models, in order to calibrate our model \citep{talts_validating_2018}.
In each case we used the 10\% of reserved simulation in the test set.
We produced and considered P-P plots, to compare the CDF of the probability integral transform to that of a uniform random variable, as well as the posterior predictions directly, to assess the accuracy and precision of each marginal posterior.
We also used approximate methods to test our posterior coverage, including the Tests of Accuracy with
Random Points \citep{lemos_robust_2023}.
TARP evaluates distances between samples of the posterior, the true value, and a random point, and uses these to estimate the expected coverage probability; this approach has been shown to provide a robust test of the optimality of the posterior estimator.
An example for the IllustrisTNG model (including LFs and colours) is shown in \fig{TNG_TARP}, showing excellent coverage and little to no bias.
Similar results are achieved across all of our trained models.

\section{Constraints from the intrinsic emission}
\label{sec:intrinsic}

\begin{figure}
    \includegraphics[width=\columnwidth]{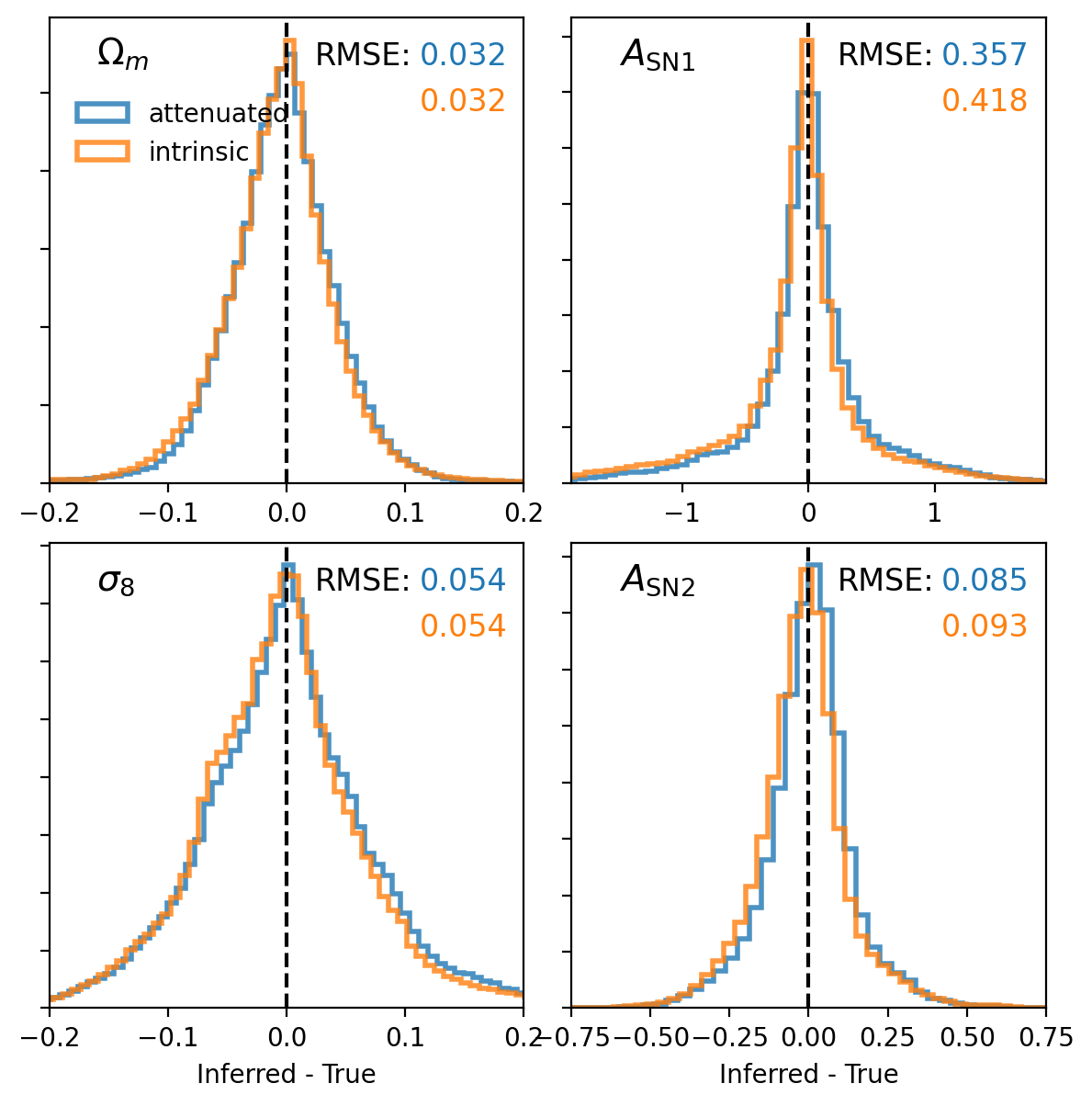}
    \caption{Combined marginal posterior distribution of \omegam, \sigmae, \asnone and \asntwo\ from all test set objects in the \textsc{IllustrisTNG} suite for the \textit{intrinsic} (orange) and \textit{dust attenuated} (blue) LFs and colours.}
    \label{fig:feature_importance_intrinsic}
\end{figure}

The dust model described in \sec{dust_model} does not depend on galaxy properties explicitly.
However, it does attenuate young stars (< 10 Myr) more than old stars, and as such it depends on the shape of the star formation history of each galaxy around this pivot point.
Galaxies with younger star formation histories will experience more attenuation; this may then act either as a degeneracy with the underlying cosmological and astrophysical parameters, or provide additional information.
To test this, we present in \fig{feature_importance_intrinsic} the stacked marginal posterior distributions using attenuated or intrinsic LFs and colours in the \textsc{IllustrisTNG} model. 
For the cosmological parameters there is no difference in the RMSE, however for the supernovae feedback parameters the constraints are slightly tighter in the model using attenuated emission compared to the intrinsic emission.
This effect, while small, suggests that the differential attenuation of old and young stars imprints additional information from the feedback parameters on the emission, from which the model can learn.

\section{Feature importance across additional models}
\label{sec:feature_importance_models}

\begin{figure*}
    \includegraphics[width=\textwidth]{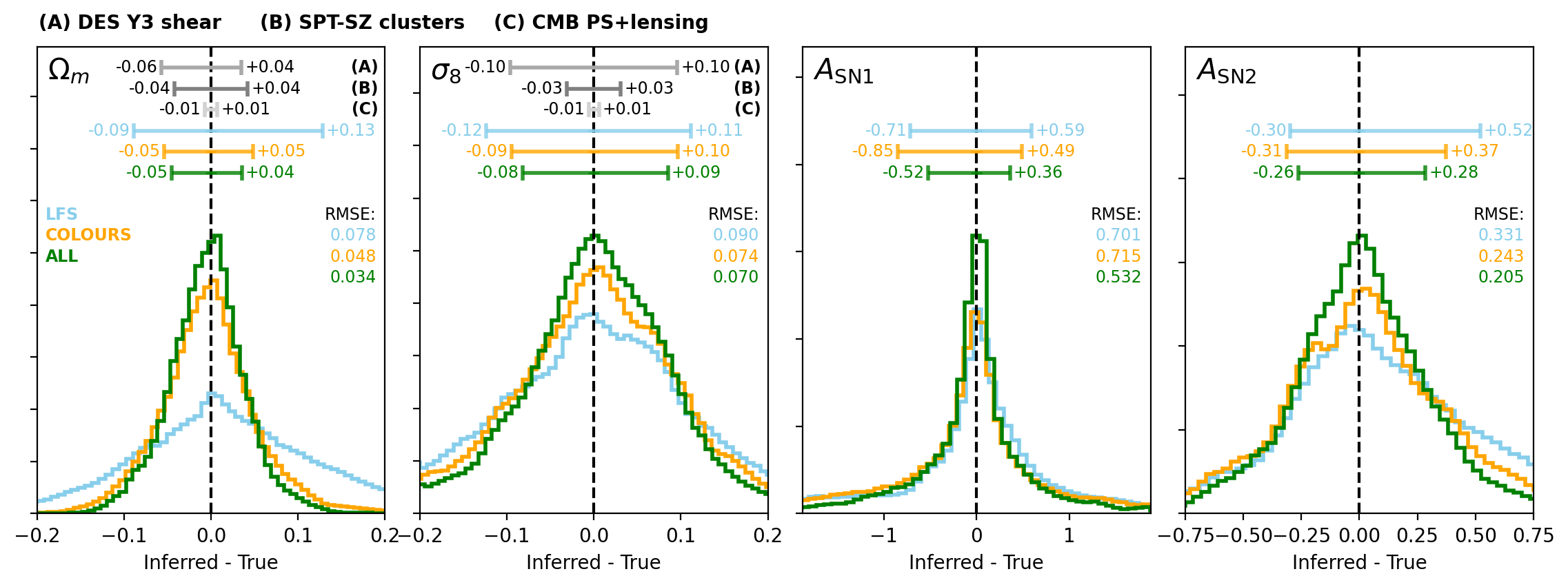}
    \includegraphics[width=\textwidth]{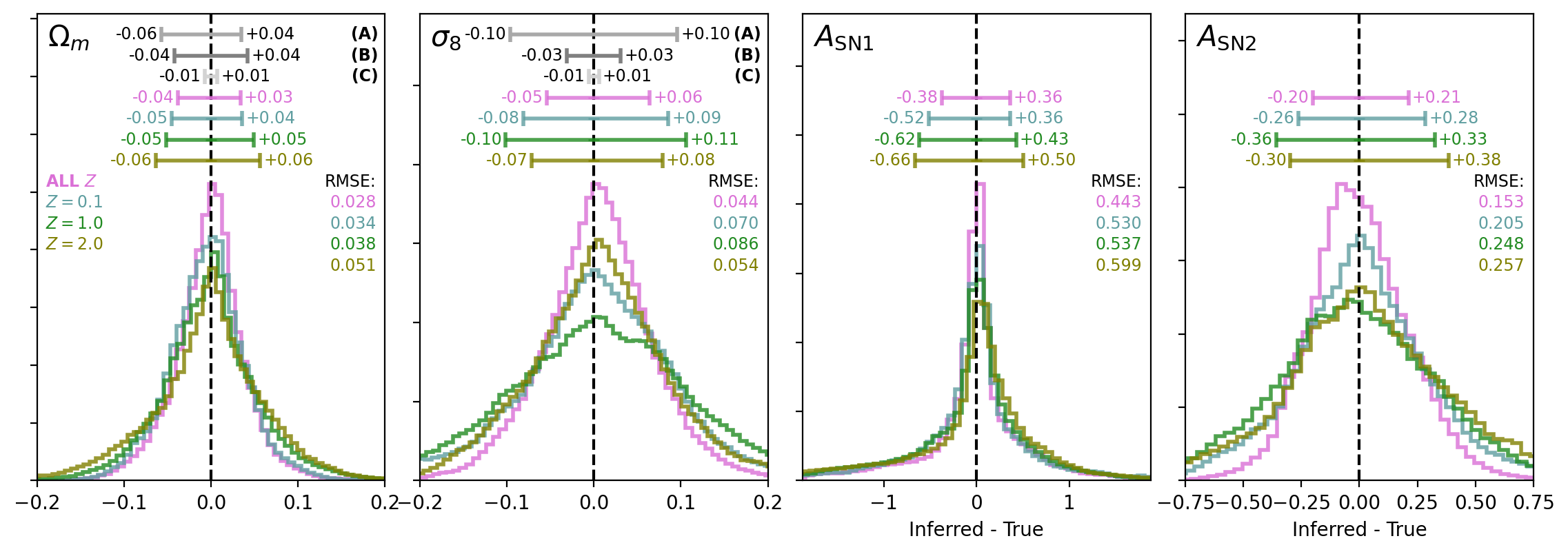}
    \caption{The same as \fig{feature_importance}, but showing the combined posteriors for the \textsc{Simba} simulations.}
    \label{fig:feature_importance_simba}
\end{figure*}

\begin{figure*}
    \includegraphics[width=\textwidth]{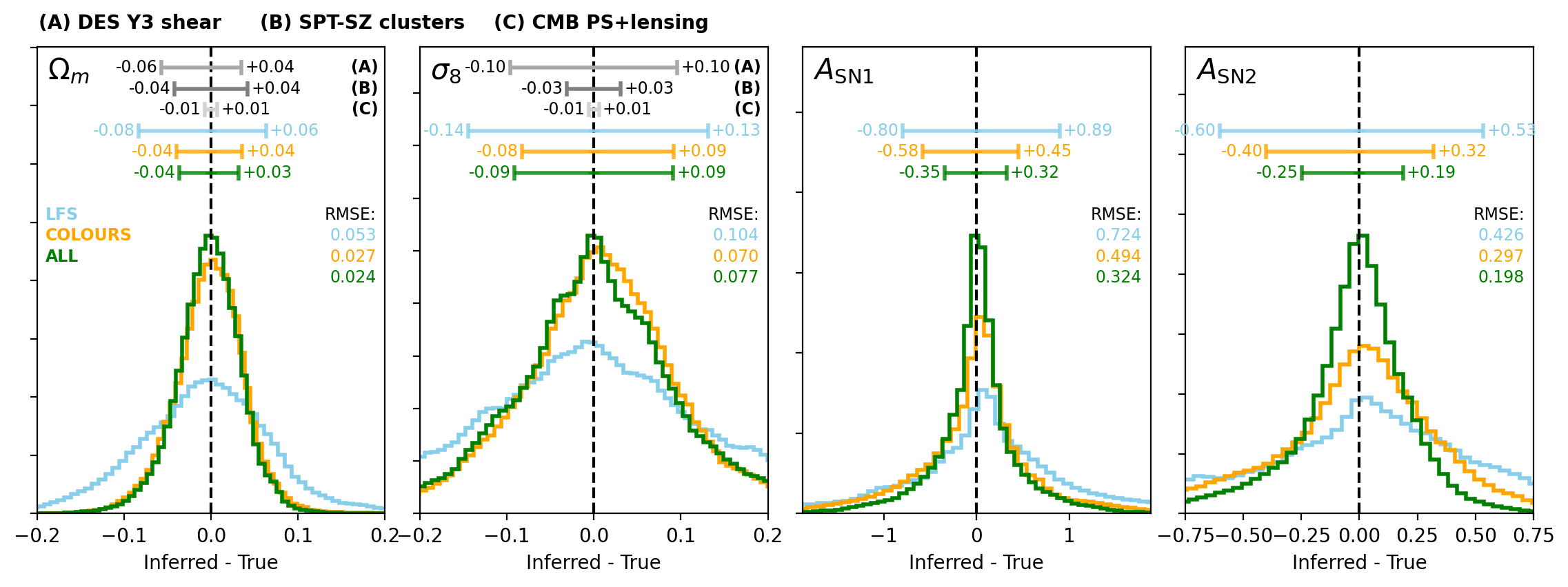}
    \includegraphics[width=\textwidth]{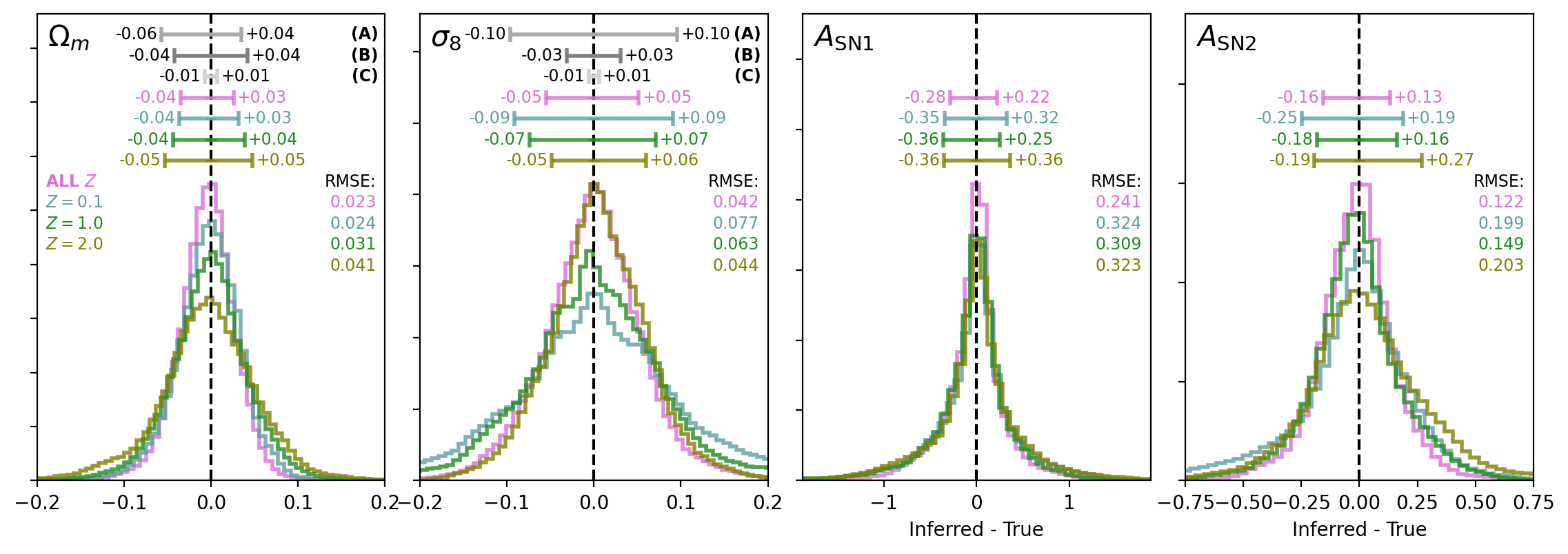}
    \caption{The same as Figures \ref{fig:feature_importance} and \ref{fig:feature_importance_simba}, but showing the combined posteriors for the \textsc{Swift-EAGLE} simulations.}
    \label{fig:feature_importance_eagle}
\end{figure*}

\begin{figure*}
    \includegraphics[width=\textwidth]{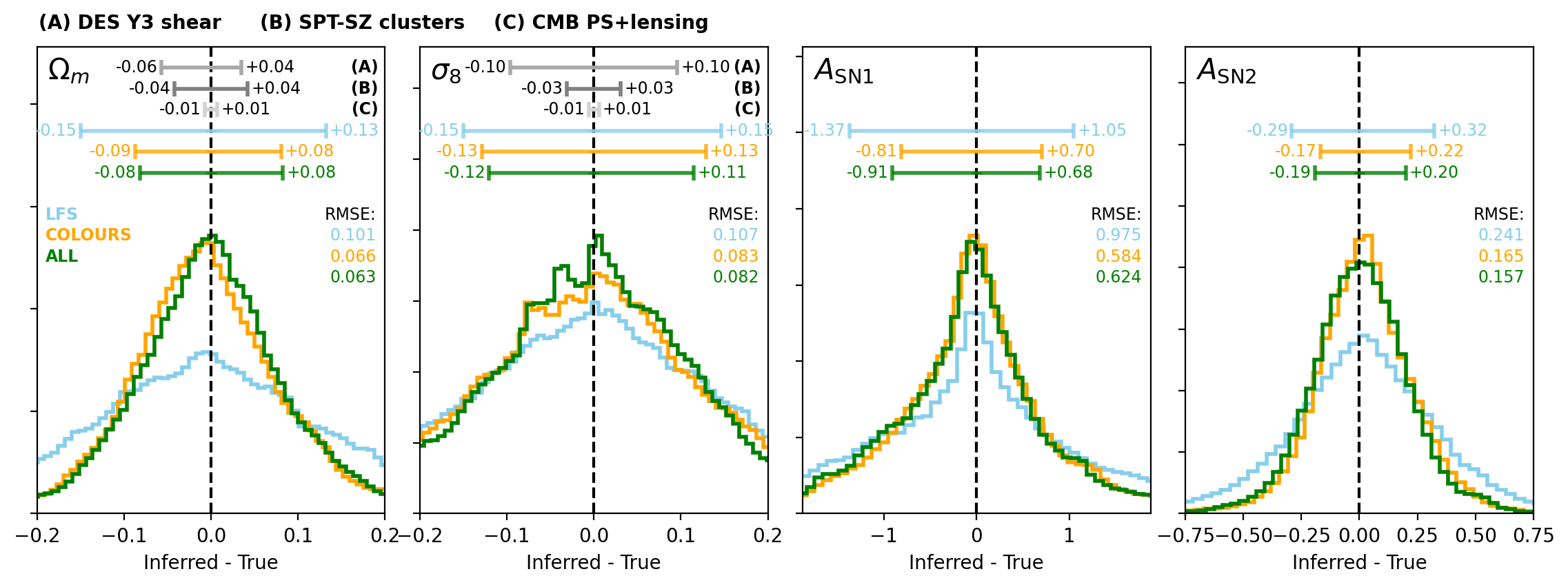}
    \includegraphics[width=\textwidth]{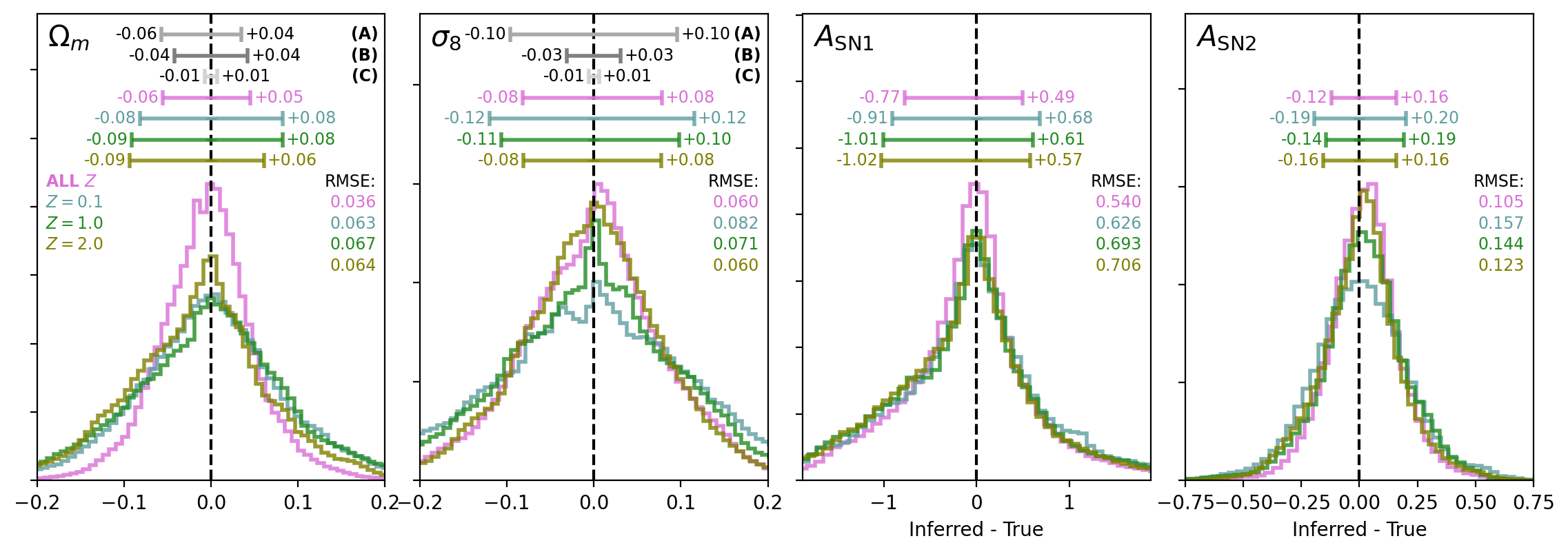}
    \caption{The same as Figures \ref{fig:feature_importance} and \ref{fig:feature_importance_simba}, but showing the combined posteriors for the \textsc{Astrid} simulations.}
    \label{fig:feature_importance_astrid}
\end{figure*}

Figures \ref{fig:feature_importance_simba}, \ref{fig:feature_importance_eagle} and \ref{fig:feature_importance_astrid} show the stacked binned residuals of the marginal posteriors for the cosmological and supernovae feedback parameters in the \textsc{Simba}, \textsc{Swift-EAGLE} and \textsc{Astrid} simulations.
Similarly to the trends seen in \fig{feature_importance} for \textsc{IllustrisTNG}, we find that for all simulations it is the colours that provide the most information on \omegam.
We also see that the information at $z = 2$ is most informative for constraints on \sigmae, as compared to lower redshift luminosity functions and colours.

\bsp	
\label{lastpage}
\end{document}